\documentclass[aps,twocolumn,prl,superscriptaddress,amsmath,amssymb,amsfonts]{revtex4-1} 

\usepackage[T1]{fontenc}
\usepackage[latin9]{inputenc}
\usepackage{lmodern} 
\usepackage{color}
\usepackage{bbold}
\usepackage{dsfont}
\usepackage{graphicx}
\usepackage[caption=false]{subfig}
\usepackage[colorlinks]{hyperref}
\usepackage{physics}
\usepackage{amsmath}
\usepackage{amssymb}
\usepackage{array} 

\begin{document}

\title{On-demand indistinguishable single photons from an efficient and pure source based on a Rydberg ensemble}

\author{D.P. Ornelas-Huerta}
\affiliation{Joint Quantum Institute, NIST/University of Maryland, College Park, MD 20742, USA}
\author{A.N. Craddock}
\affiliation{Joint Quantum Institute, NIST/University of Maryland, College Park, MD 20742, USA}
\author{E.A. Goldschmidt}
\affiliation{Department of Physics, University of Illinois at Urbana-Champaign, 1110 W. Green Street, Urbana, Illinois 61801, USA}
\affiliation{US Army Research Laboratory, Adelphi, Maryland 20783, USA}
\author{A.J. Hachtel}
\affiliation{Joint Quantum Institute, NIST/University of Maryland, College Park, MD 20742, USA}
\author{Y. Wang}
\affiliation{Joint Quantum Institute, NIST/University of Maryland, College Park, MD 20742, USA}
\author{P. Bienias}
\affiliation{Joint Quantum Institute, NIST/University of Maryland, College Park, MD 20742, USA}
\affiliation{Joint Center for Quantum Information and Computer Science, NIST/University of Maryland, College Park, MD 20742, USA}
\author{A.V. Gorshkov}
\affiliation{Joint Quantum Institute, NIST/University of Maryland, College Park, MD 20742, USA}
\affiliation{Joint Center for Quantum Information and Computer Science, NIST/University of Maryland, College Park, MD 20742, USA}
\author{S.L. Rolston}
\affiliation{Joint Quantum Institute, NIST/University of Maryland, College Park, MD 20742, USA}
\author{J.V. Porto}
\affiliation{Joint Quantum Institute, NIST/University of Maryland, College Park, MD 20742, USA}
\email{Corresponding author: porto@umd.edu}

\begin{abstract}
Single photons coupled to atomic systems have shown to be a promising platform for developing quantum technologies. Yet a bright on-demand, highly pure and highly indistinguishable single-photon source compatible with atomic platforms is lacking. In this work, we demonstrate such a source based on a strongly interacting Rydberg system. The large optical nonlinearities in a blockaded Rydberg ensemble convert coherent light into a single-collective excitation that can be coherently retrieved as a quantum field. We observe a single-transverse-mode efficiency up to 0.18(2), $g^{(2)}=2.0(1.5)\times10^{-4}$, and indistinguishability of 0.982(7), making this system promising for scalable quantum information applications. Accounting for losses, we infer a generation probability up to 0.40(4). Furthermore, we investigate the effects of contaminant Rydberg excitations on the source efficiency. Finally, we introduce metrics to benchmark the performance of on-demand single-photon sources.
\end{abstract}

\maketitle

\section*{Introduction}

Engineering single-photon sources with high efficiency, purity, and indistinguishability is a longstanding goal for applications such as linear optical quantum computation \cite{carolan2015}, boson sampling \cite{wang2019}, quantum networks \cite{yin2017} and quantum metrology \cite{slussarenko2017}. Atomic systems have shown significant progress towards quantum light-matter interfaces, including efficient quantum memories \cite{wang2019efficient}, quantum networks \cite{yu2020}, high-fidelity light-matter entanglement \cite{bock2018}, atomic gates \cite{ballance2016}, and quantum simulators \cite{gross2017}. Atomic platforms require spectrally matched single photons that can coherently couple with atomic processors, provided with high-efficiency generation, purity, and indistinguishability. 

Strongly interacting Rydberg atoms provide a particularly promising system. They have proven to be versatile for engineering strong interactions between photons, exhibiting nonlinearities at the single-photon level \cite{peyronel2012, maxwell2013, li2016, paris2017}. Recent experiments using Rydberg interactions have demonstrated on-demand single-photon generation \cite{dudin2012, ripka2018}, as well as photon transistors \cite{gorni2014, tiarks2014, gorni2016}, photonic and atomic phase gates \cite{tiarks2016, thompson2017,tiarks2018, maller2015, zeng2017, levine2018}, high-visibility quantum interference in hybrid systems \cite{us}, and quantum simulators \cite{schauss2012, zeiher2017,lienhard2018, kim2018}.

We describe here an efficient single-photon source based on collective excitation and de-excitation of a cold, trapped ensemble of atoms through a highly excited Rydberg state \cite{saffman2002, dudin2012, ripka2018}. During two-photon excitation from the ground to the Rydberg state via an intermediate state [see Fig.~\ref{fig:exp}(a)], long-range van der Waals interactions suppress multiple Rydberg excitations within a blockade radius, $r_b$ \cite{lukin2001}. The resulting single, collective atomic excitation is coherently shared among $N$ atoms as a spin wave \cite{saffman2002}. Due to the collective nature of the excitation, if the initial phase coherence of the spin wave is maintained, the subsequent coupling of the Rydberg state to the intermediate state can efficiently map the excitation onto a single photon in a well-defined mode~\cite{sangouard2011}. Our system produces single photons with repetition rates up to 400~kHz, a generation probability up to 0.40(4), $g^{(2)}=2.0(1.5)\times10^{-4}$,
and indistinguishability of 0.982(7). We model the write and retrieval process, including the measured spin-wave dephasing rate. We identify long-lived-contaminant Rydberg states \cite{Elizabeth2016} as a limiting factor on the source efficiency for increasing production rates.

Given the requirements for most quantum information applications, the single-mode efficiency, rate, and quality of single-photon sources are of key importance since successful scaling of these systems involves detection of multiple identical photons. Thus, we introduce metrics to describe the probability, rate, and fidelity of producing a single photon in a single-mode, which includes the contributions from the commonly used metrics: overall collection efficiency, purity, indistinguishability, and repetition rate \cite{eisaman2011}.

\section*{Experimental Apparatus and Procedure } 

\begin{figure}[t]
\centering\includegraphics[width=8.4 cm]{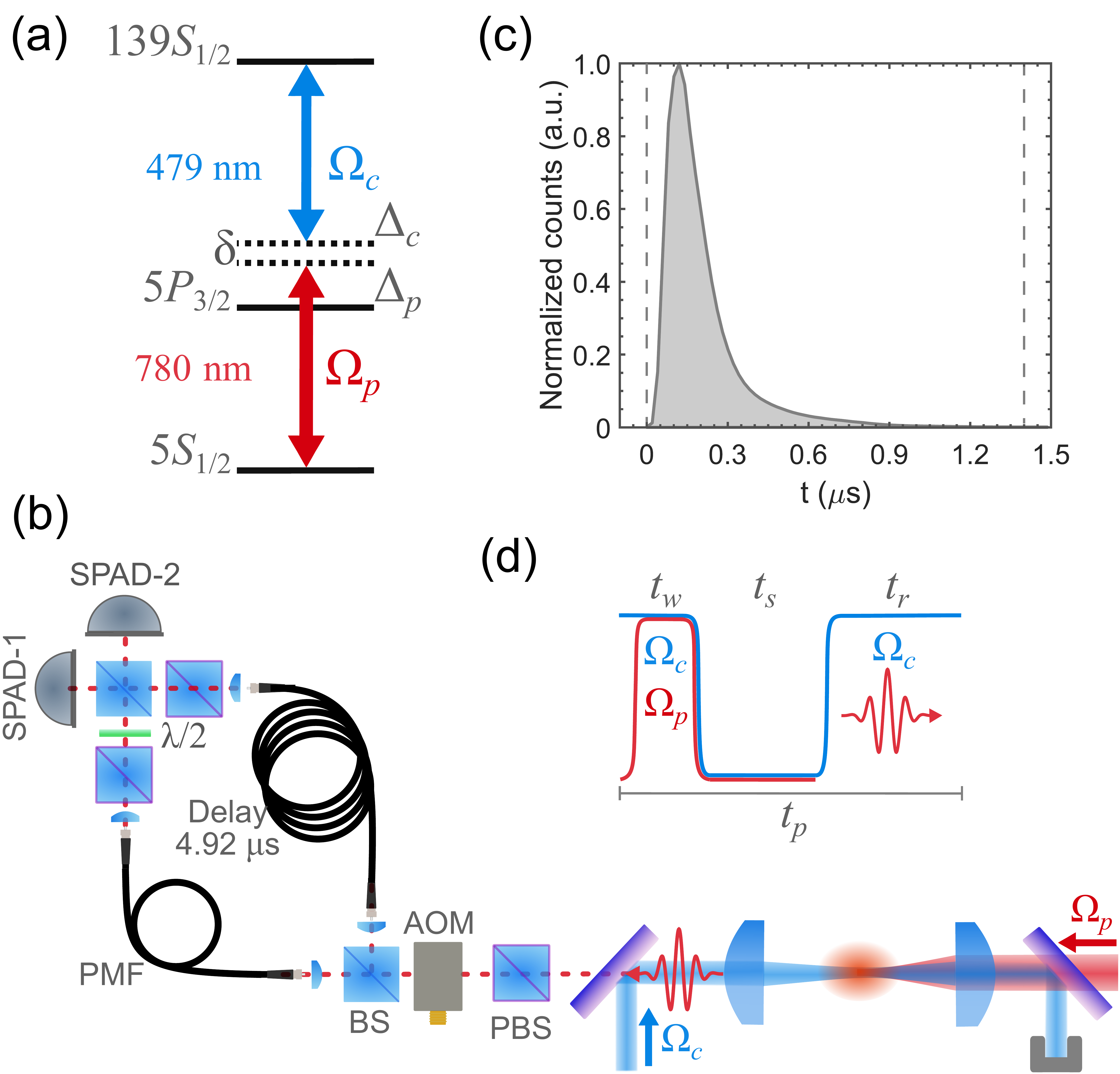}
\caption{(a) Relevant atomic levels and set-up for single-photon generation. During the spin wave writing stage we set the single-photon detuning  $\Delta_p\approx2\pi\times50$~MHz, and the two-photon detuning $\delta=\Delta_p+\Delta_c$ to Raman resonance, $\delta\approx-2\pi\times2$~MHz. For retrieval,  $\Delta_c\approx2\pi\times7$~MHz. (b) Experimental set-up schematic. There is a polarization beamsplitter (PBS) to project the photons into a single polarization mode, followed by an acousto-optic-modulator (AOM) that gates the incoming photons. All the light is directed to the polarization maintaining fiber (PMF) to realize a purity measurement. For the indistinguishability characterization, we split the light such that the rate is roughly the same at both ports of the second beamsplitter (BS). By rotating the half waveplate ($\lambda/2$) we can control the relative polarization of the photons coming from the PMF port and the long delay port. (c) Photon temporal envelope, gray dashed lines indicate the software gate window. (d) Timing sequence for the generation of successive single photons, the writing $\pi$-pulse lasts for $t_w\approx370$~ns. We use a minimum storage time $t_s\approx350$~ns to maximize the retrieval and vary $t_r$ to change the repetition rate, $R=1/t_p$. }
\label{fig:exp}
\end{figure}

We start the experiment with a magneto-optical trap of $^{87}$Rb atoms and further laser cool the atoms with a $\Lambda$-gray molasses down to $\approx10$ $\mu$K. We load the atoms into a 1003-nm wavelength optical dipole trap. To write the spin wave, we couple the ground state, $|g\rangle = |5S_{1/2}, F=2, m_F=2\rangle$ to the Rydberg state $|r\rangle =|139S_{1/2}, m_J=1/2\rangle$ via the intermediate state $|e\rangle= |5P_{3/2}, F=3, m_F=3\rangle$ with an intermediate detuning $\Delta_p\approx2\pi\times50$~MHz, as shown in Figure \ref{fig:exp}(a). The probe beam coupling $|g\rangle$ to $|e\rangle$  is focused into the atom cloud with a waist of $\approx3.3$~$\mu$m, with a Rabi frequency $\Omega_p \approx 2 \pi \times 1$~MHz. The counter-propagating control beam coupling $|e\rangle$ to $|r\rangle$ has a larger, $\approx19$~$\mu$m waist and peak Rabi frequency $\Omega_c \approx 2 \pi \times 7$~MHz.

The van der Waals coefficient of the Rydberg state $139S_{1/2}$ is $C_6 \approx -2 \pi \times 2.5\times10^6$~GHz~$\mu$m$^6$ \cite{ARC}, which results in a blockade radius $r_b \approx 60$~$\mu$m during the spin-wave writing.  Since $r_b$ is larger than the probe beam waist and the atomic cloud extension in the propagation direction, $\sigma_z\approx27$ $\mu m$, the excitation volume is blockaded. The effective two-photon Rabi frequency, $\Omega_{\text{2ph}}= \frac{\Omega_p \Omega_c }{2\Delta_p}$ is enhanced by a factor $\sqrt{N}\approx20$ from the $N$ atoms participating in the collective excitation \cite{saffman2002, dudin2012coll}.

After a spin-wave storage time $t_s > 350$~ns [see Fig.~\ref{fig:exp}(d)], we turn back on the control field with a detuning $\Delta_c \approx 2 \pi \times 7$~MHz that maximizes the retrieval efficiency of the spin wave into a single photon. We can vary the repetition rate of the write-retrieval sequence up to 400~kHz, with interrogation times up to 600~ms (0.6 duty cycle) before we need to reload the optical dipole trap.

\section*{Single-photon source purity and indistinguishability}

\begin{figure}
\centering\includegraphics[width=7.5cm]{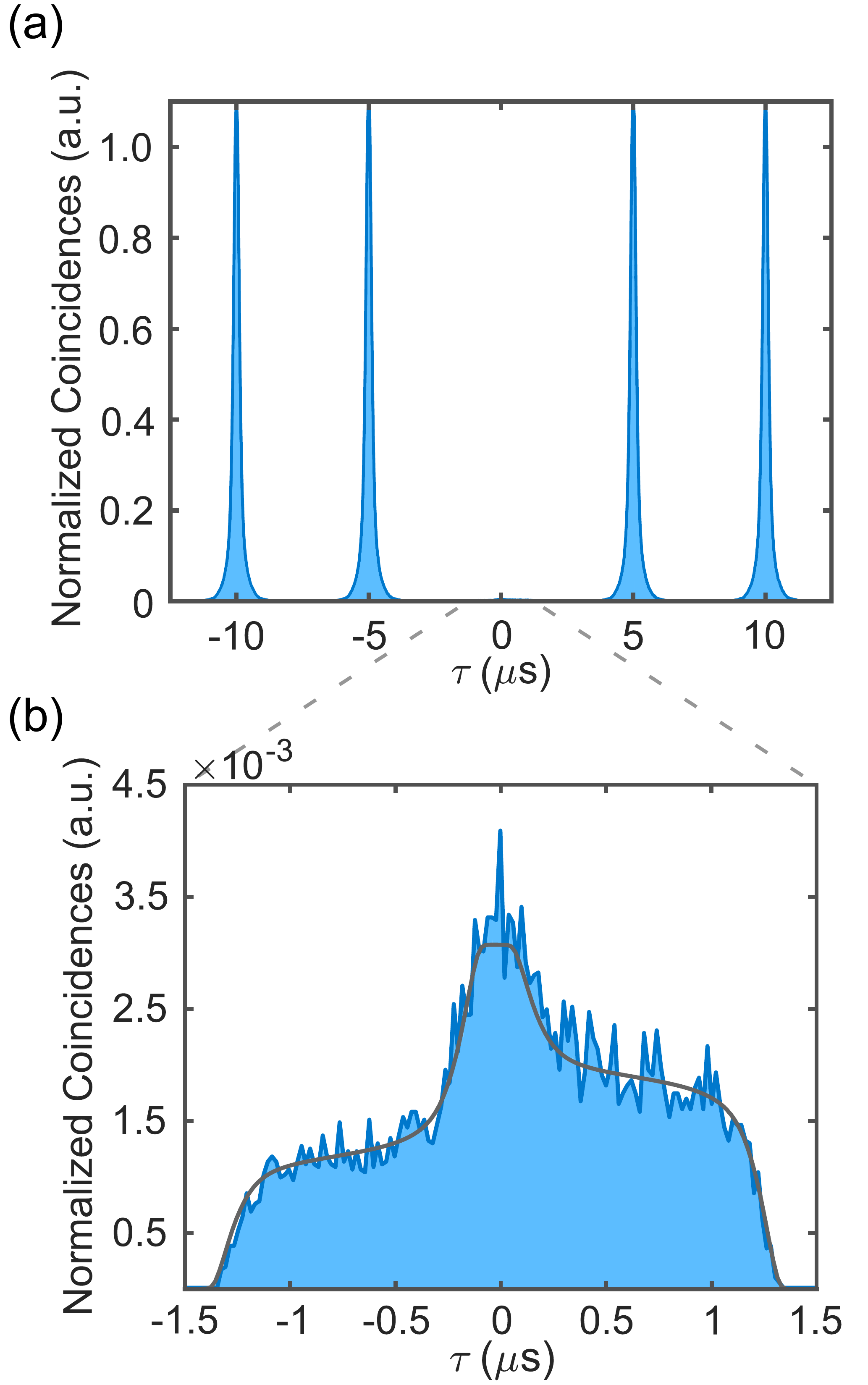}
\caption{Measured coincidences for purity and characterization. (a) Normalized coincidences for $g^{(2)}(\tau)$ with 5 $\mu$s cycle. (b) Normalized coincidences for $g^{(2)}(\tau)$ around $\tau=0$, grey line represents the background coincidences with 20-ns bins. The shape of this profile arises from the convolution of the photon pulse shape with a constant background within the gate window, and the pedestal asymmetry is because the background rate is not the same for each channel. All data shown were taken with $60\%$ duty cycle.}
\label{fig:g2}
\end{figure}

We use Hanbury Brown-Twiss and Hong-Ou-Mandel interferometers to characterize the purity and indistinguishability of our single photons [see Fig.\ \ref{fig:exp}(b)]. We define the purity of our single-photon source as $1-g^{(2)}(0)$, where $g^{(2)}(\tau)$ is the second-order autocorrelation function. We apply a 1.4 $\mu$s long software gate window, containing more than $99.9\%$ of the pulse [see Fig.\ \ref{fig:exp}(c)]. Coincidences at zero time delay are substantially suppressed, as shown in Figure \ref{fig:g2}(a), with strong antibunching $g^{(2)}_{\text{raw}}(0)=0.0145(2)$, integrating the area around $\tau=0$ and without background subtraction. The background coincidence rate is dominated by coincidences involving photon events with background counts unrelated to the single-photon generation, coming from detector dark counts and room light leakage. The independently measured background rate, photon shape, and photon rate are constant throughout each experimental run, from which we determine that the accidental coincidences contribute to $g^{(2)}_{\text{back}}(0)=0.0143$. The gray curve in Figure \ref{fig:g2}(b) shows the background coincidence profile within the gate window (see \cite{SM} for details). After background subtraction, our single-photon source has $g^{(2)}(0)=2.0(1.5)\times 10^{-4}$.

\begin{figure}
\centering\includegraphics[width=7.5cm]{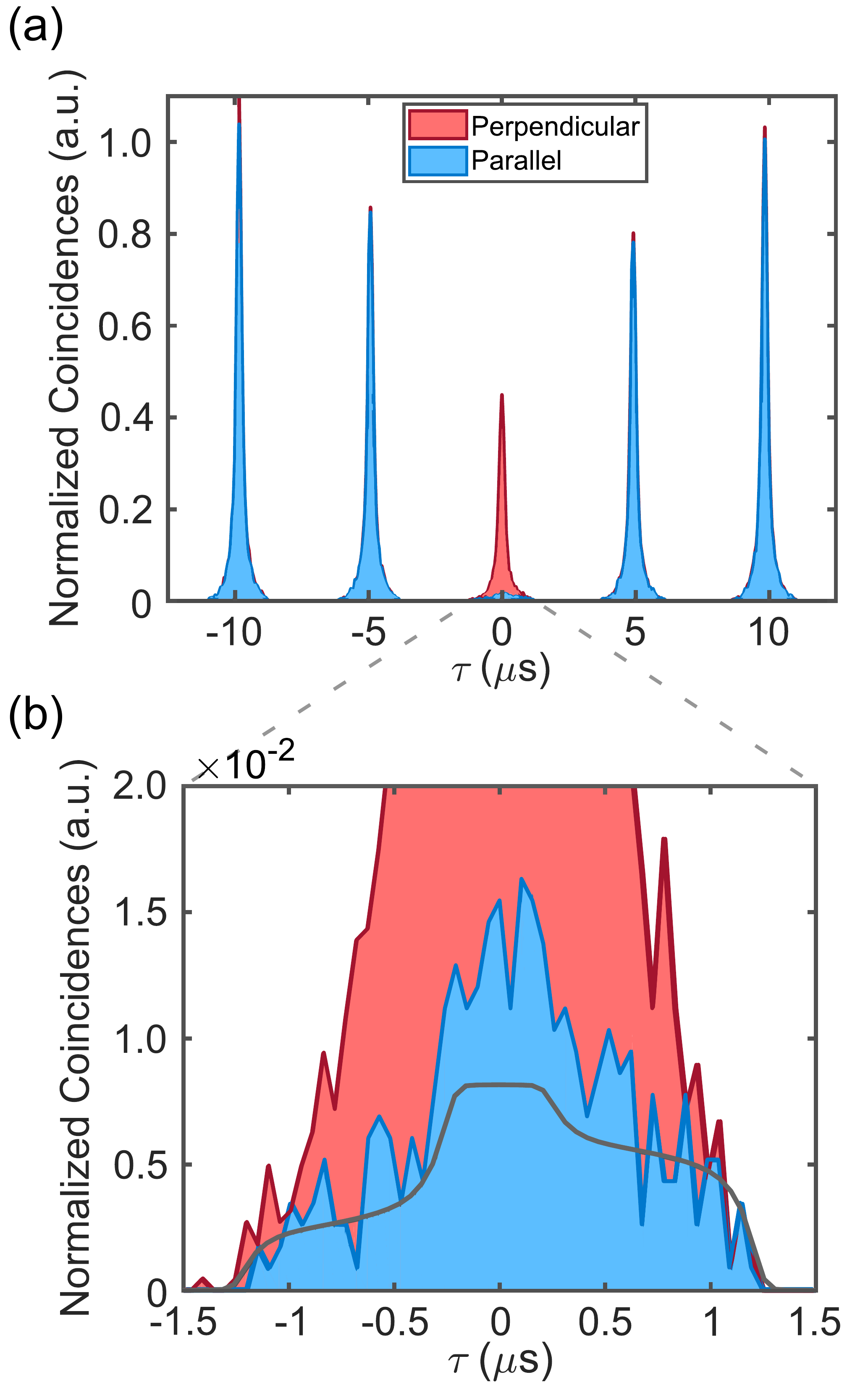}
\caption{Measured coincidences for indistinguishability characterization. (a) Normalized coincidences for HOM characterization with 4.92 $\mu$s cycle. Indistinguishable polarization states are represented in blue, and distinguishable polarization states are in red. (b) Normalized coincidences for HOM around $\tau=0$, the grey line represents the background coincidences with 52-ns bins. All data shown were taken with $60\%$ duty cycle.}
\label{fig:HOM}
\end{figure}

We use a Hong-Ou-Mandel interferometer (HOM) to measure the photon indistinguishability. We implement a fiber-based $4.92~\mu$s delay in one arm to temporally overlap adjacently produced photons. Additionally, there is a polarizing beam splitter (PBS) at the output of each fiber to account for any polarization rotation due to the fibers. At the exit of the short arm, there is a half-wave plate (HWP) to rotate the polarization and control the degree of distinguishability of the photons. Figure \ref{fig:HOM}(a) shows the normalized coincidences for orthogonal and parallel polarizations. Integrating the number of coincidences in a window around $\tau=0$ for the two cases, we measure a raw HOM interference visibility $\mathcal{V}_{\text{raw}}=1-C_{\parallel}/C_{\perp}=0.894(6)$. 
Accounting for the accidental coincidences with background events and the slight differences in the transmission and reflection coefficients of our combining beamsplitter gives a mode overlap of 0.982(7) (see \cite{SM}).

\section*{Source efficiency}

\begin{figure*}[t]
\centering\includegraphics[width=0.75\textwidth]{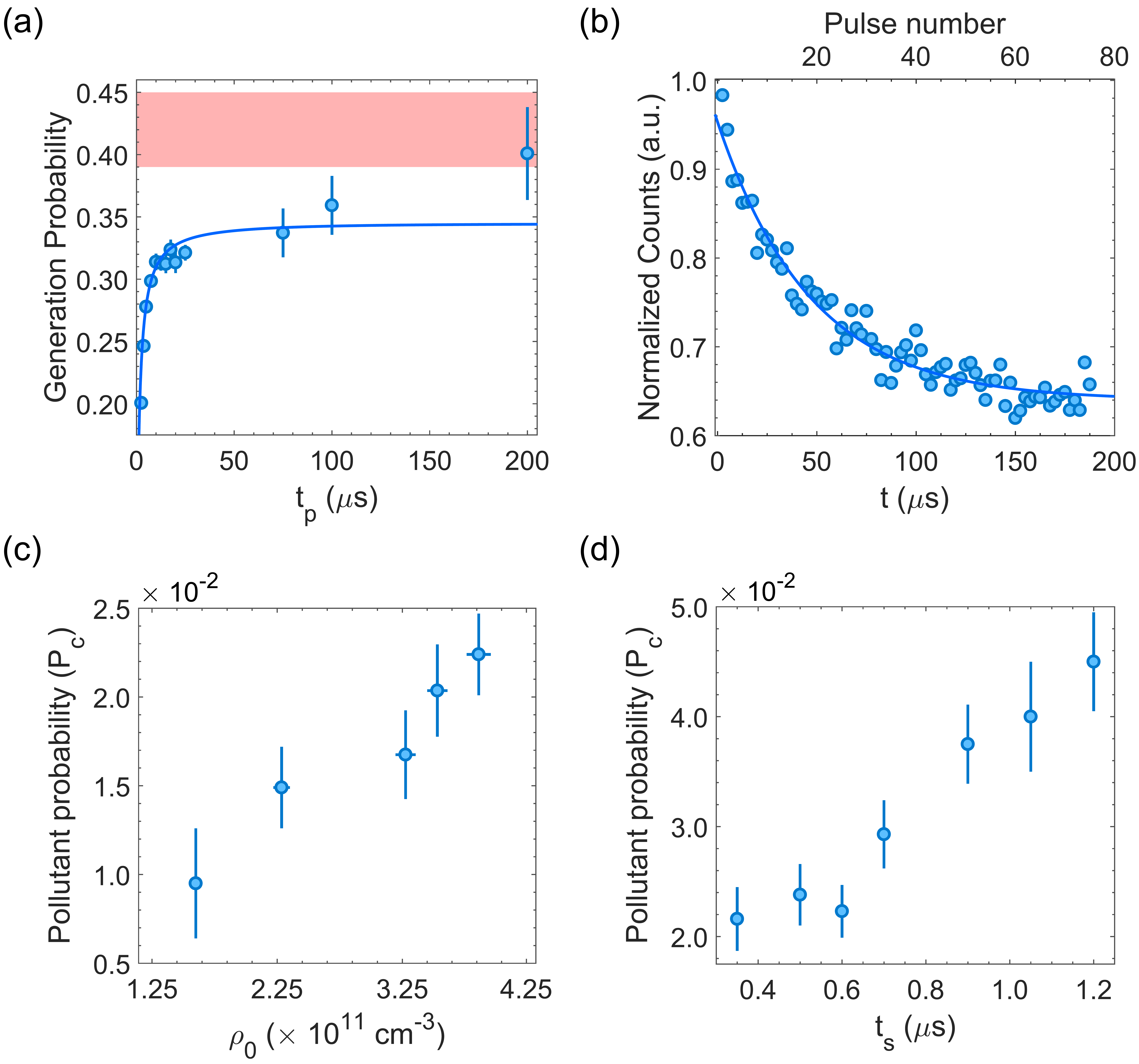}
\caption{Effect of contaminants on single-photon generation. (a) Photon generation probability as function of pulse period $t_p$. Dark-blue line is fitted using Eq.~\ref{eq:prob} in steady state for $n \rightarrow\infty$ using the values for $P_c$ and $\tau_c$ in the main text, we obtain $P_{max}=0.35(2)$. Red band shows the generation probability predicted by the theoretical model.  (b) Normalized summed counts per pulse for a pulse train with 2.5-$\mu$s pulse period. Dark-blue line is fitted with Eq.~\ref{eq:prob}. (c) $P_c$ vs. peak atomic density $\rho_0$ with a fixed storage $t_s=350$~ns.  (d) $P_c$ vs.  time $t_s$ with a density  $\approx4\times10^{11}$~cm$^{-3}$.}
\label{fig:contaminants}
\end{figure*}

We measure a peak probability of 0.18(2) to generate a single photon into a single-mode fiber after polarization filtering and averaged for a 20$\%$ duty cycle. Accounting for optical losses and assuming that the single-photon has the same spatial mode as the 780-nm-write beam, we estimate a  generation probability of 0.40(4) immediately after the atomic ensemble. The average probabilities go down to 0.14(1) and 0.31(1), respectively for a $60\%$ duty cycle.

We calculate $P_{\text{th}}=\eta_w \eta_s \eta_r$  as a product of the writing, $\eta_w$, storage, $\eta_s$, and retrieval, $\eta_r$, efficiencies  to estimate the theoretical probability of generating a photon. Referring the reader to the Supplement \cite{SM} for the details of the theoretical analysis, we summarize it here only briefly. We simulate the writing of the spin wave using a Lindblad master equation to estimate the writing efficiency and the storage efficiency. We calculate the retrieval efficiency using the optical Maxwell-Bloch equations with the formalism in Ref.\ \cite{alexey2007}. Using independently measured experimental values as input parameters, we obtain a theoretical prediction of $P_{\text{th}}\approx0.42(3)$ (see Supplement \cite{SM}). This value is consistent with the measured generation probability for the longest pulsing periods, $t_p$.

We observed that the average photon production efficiency decreased at higher repetition rates, as shown in Figure~\ref{fig:contaminants}(a). (Here the photon probability is determined immediately after the atom cloud by accounting for independently measured optical losses.) The initial pulse in a pulse series had higher efficiency, however, the efficiency of subsequent pulses decreased exponentially to the steady-state value on a $\approx60$~$\mu$s time scale [see Figure~\ref{fig:contaminants}(b)]. 

These observations are consistent with the creation of contaminant atoms in other long-lived Rydberg states that are not removed by the retrieval field. These states interact strongly with the target Rydberg state, affecting subsequent writing events. Similar contaminant states have been observed in previous experiments \cite{DeSalvo2016, Elizabeth2016, Radiation2017}, and have been analyzed extensively \cite{Aman2016, Chem2016, boulier2017, bienias2018}. Once a contaminant is in the medium, it disables the writing of a spin wave for the later pulses. However, contaminants have a finite lifetime in the medium, therefore, the photon generation probability decreases for shorter pulse periods.

We use a simple model to capture the effect of contaminants on photon production (see \cite{SM} for details). We assume that for any given pulse, there is a probability $P_c$ of creating a contaminant. If the contaminant state has a lifetime $\tau_c$, then the probability $P_n$ of having a contaminant in the $n$-th pulse of a pulse series with period $t_p$ is
\begin{equation}
    P_n=P_c \frac{1-(e^{- t_p/\tau_c}-P_c)^n  }{1-e^{-t_p/\tau_c}+P_c}.
    \label{eq:prob}
\end{equation}
For $\tau_c \gg t_p$, the average contaminant probability as $n\rightarrow \infty$ can be significant, even if $P_c$ is small. 
The probability $P_g(n)$ of successfully generating a single-photon on the $n$-th pulse in the presence of a contaminant is decreased according to $P_g(n)=P_{max}(1-P_n)$, where $P_{max}$ is the probability of photon generation in the absence of contaminants. The steady state efficiency is given by $P_g(n\rightarrow\infty)$. 
Fitting this equation to pulse sequence data as shown in Fig.~\ref{fig:contaminants}(b), we determine $P_c=1.9(3)\times10^{-2}$, and $\tau_c=65(8)$~$\mu$s, which is in good agreement with the data in Fig.~\ref{fig:contaminants}(a).

We find that $P_c$ increases linearly with atomic density $\rho$ [see Fig.~\ref{fig:contaminants}(c)], which suggests that the source of contaminants is ground-Rydberg interactions. For high principal quantum number, $n$, collisionally produced contaminants were identified in Ref~\cite{Chem2016} to be Rydberg states with principal quantum number $n-4$ and quantum angular momentum $l>2$. Furthermore, we find that $P_c$ increases with storage time $t_s$ at a rate $\approx3\times10^{-2}$~$\mu$s$^{-1}$, which gives a contaminant generation time-scale of $\approx33$~$\mu$s for a density $\approx4\times10^{11}$~cm$^{-3}$. Contaminants are not a fundamental limitation since strong electric field pulses between writing pulses could be used to remove them.

We also note that for interrogation times longer than 100~ms, other effects such as heating and atom depolarization from rescattering become more significant, further reducing the photon generation for shorter $t_p$. However, these effects can be mitigated by detuning farther from the intermediate state.

\begin{figure}
\centering\includegraphics[width=8cm]{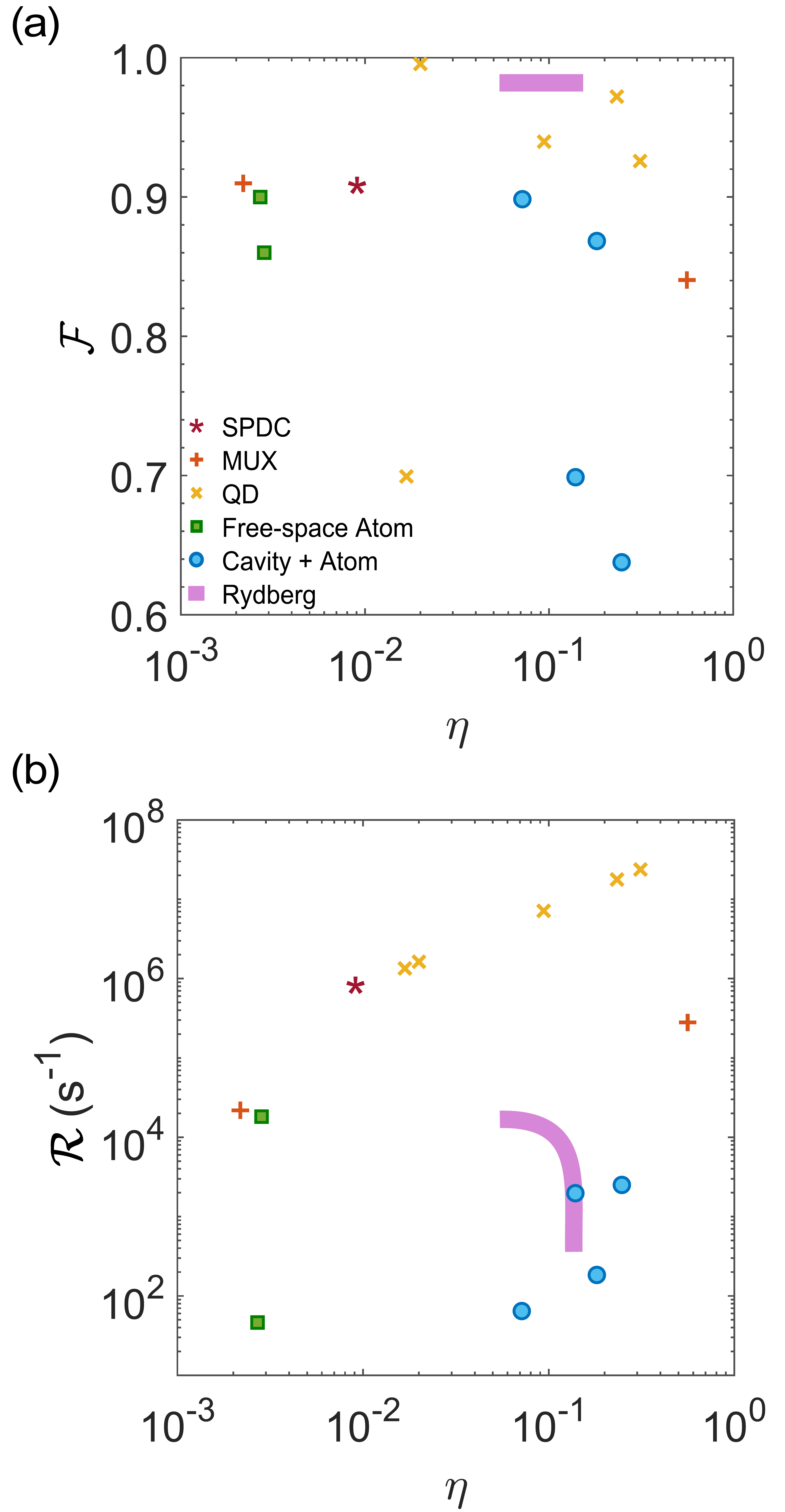}
\caption{Performance of a sample from different single-photon sources. Solid-state systems considered are spontaneous parametric down-conversion (SPDC) \cite{wang2016}, multiplexed-heralded-single-photon source (MUX-HSPS) \cite{xiong2016, kaneda2019} and quantum dots (QD) \cite{somaschi2016, loredo2016, wang2017, kir2017, wang2019p}. Atomic systems considered are single atoms in free-space \cite{maunz2007, rosenfeld2017}, atoms in cavities \cite{thompson2006, wilk2007, nisbet2011, mucke2013}, and the Rydberg ensemble studied in this work (indicated in the purple line) accounting for the effect of different repetition rates for a duty cycle of 0.6. (For details on these sources, see tables in  \cite{SM}). (a) Fidelity vs. single-mode efficiency. (b) Brightness vs. single-mode efficiency.}
\label{fig:comparison}
\end{figure}

\section*{Single-mode efficiency, rate and fidelity}

There are many metrics used to quantify the various properties of single-photon sources. Optical quantum information schemes are susceptible to errors if they are not implemented with highly pure and indistinguishable single photons. In addition, scaling up quantum information protocols needs high generation efficiency, since any inefficiency will  lead to an exponential decrease of the success probability with system size. Finally, the rate of single-photon production provides a limitation on the practicality of any protocol. To that end, we define three metrics that quantify these properties: $\mathcal{F}$, the single-photon fidelity, which is the fraction of emission that consists of a single photon in a single spectral, temporal, polarization, and spatial mode; $\eta$, the probability of generating a single photon in the desired mode; and $\mathcal{R}$, the brightness, which the rate of photon production in the desired mode. 

Assuming that the probability of multi-photon events greater than two is negligible,  the only outcomes from a source are:  single photons in the desired mode with probability $\eta$, single photons in an undesirable mode with probability $P_1'$, two photons with probability $P_2$, and null events with probability $P_0$. 
Experimentally, we measure the following quantities: the overall emission efficiency, $P=1-P_0$; the HOM visibility, $\mathcal{V}$; and the measure of the single-photon purity, $g^{(2)}$. These are given by:

\begin{equation}
   \begin{split}
    P & = 1-P_0=\eta+P_1'+P_2,\\
    \mathcal{V} & = \frac{\eta}{\eta+P_1'},\\
    g^{(2)}& \approx \frac{2 P_2}{(\eta+P_1'+2P_2)^2},
    \end{split}
\end{equation}
where we have assume that the visibility $\mathcal{V}$ is compensated for multi-photon events \cite{SM}, and that these measurements are taken with standard non-number resolving photon counting detectors.

Solving the system of equation for $\eta$  to second order in $g^{(2)}$, we get the single-mode efficiency $\eta$:
\begin{equation}
    \eta=P \mathcal{V}\left(1-\frac{1}{2}P g^{(2)}\left(1+P g^{(2)}\right) \right).
    \label{eq:eta}
\end{equation}

We report the source brightness as $\mathcal{R}=R_{\text{eff}}\eta$, where $R_{\text{eff}}$, is the clock rate weighted by the experimental duty cycle. Apart from source brightness, the rate at which undesirable emission is produced also matters for applications. We characterize this rate by the fidelity, 
\begin{equation}
\mathcal{F}=1-\frac{P_1'+P_2}{P}=\frac{\eta}{P}, 
\end{equation}
which is the fraction of collected emission that is made up of single photons in the correct mode. In Fig.~\ref{fig:comparison} we show $\eta$, $\mathcal{F}$, and $\mathcal{R}$ for a sample of different single-photon sources. Narrow bandwidth sources naturally compatible with coherent atomic systems are indicated with filled symbols.

\section*{Conclusion}

By using the quantum nonlinearities of strongly interacting Rydberg states in a cold atomic ensemble, we demonstrated a single-photon source, operating with a 60$\%$ duty cycle, single-mode efficiency $\eta=0.139(5)$, a single-mode brightness of $\mathcal{R}=840(70)~s^{-1}$, and single-mode fidelity $\mathcal{F}=0.982(7)$, this fidelity is the highest reported to our knowledge for an atomic-based source. Furthermore, we investigated the limitations of our current setup arising from nearby long-lived contaminant states.
 
Implementing feasible improvements to the current experiment we estimate that we can achieve up to $\eta\approx0.4$ and moreover, ionizing pulses after each write-retrieval pulse to remove atoms in pollutant states may increase the brightness up to $\mathcal {R}\approx1.2\times10^5~s^{-1}$ without decreasing the duty cycle or the fidelity (see \cite{SM} for details). The efficiency could be further improved if the ensemble were coupled to a cavity \cite{clark2019}. Given their high efficiency, brightness, and fidelity, we have shown that single-photon sources based on Rydberg-atomic ensembles provide a promising platform for scalable quantum photonics. Furthermore, they are inherently compatible with narrow-bandwidth atomic platforms that have shown significant progress towards quantum information applications.

\section*{Acknowledgments}

All authors acknowledge support from the United States Army Research Lab's Center for Distributed Quantum Information (CDQI) at the University of Maryland and the Army Research Lab. A.C, D.O.-H, A.J.H., S.L.R., J.V.P., Y.W., P.B., and A.V.G.\ additionally acknowledge support from the National Science Foundation Physics Frontier Center at the Joint Quantum Institute (Grant No. PHY1430094). Y.W., P.B., and A.V.G.\ additionally acknowledge support from AFOSR, ARO MURI, and DoE ASCR Quantum Testbed Pathfinder program (award No. DE-SC0019040).

We are grateful to Mary Lyon for her significant contributions to the design and construction of the apparatus and Patrick Banner for his contributions to data collection. We also want to thank Luis A. Orozco for fruitful discussions.

\clearpage
\onecolumngrid

\begin{center}

\newcommand{\beginsupplement}{%
        \setcounter{table}{0}
        \renewcommand{\thetable}{S\arabic{table}}%
        \setcounter{figure}{0}
        \renewcommand{\thefigure}{S\arabic{figure}}%
     }

\textbf{\large \section{Supplemental Material}}
\end{center}

\newcommand{\beginsupplement}{%
        \setcounter{table}{0}
        \renewcommand{\thetable}{S\arabic{table}}%
        \setcounter{figure}{0}
        \renewcommand{\thefigure}{S\arabic{figure}}%
     }

\setcounter{equation}{0}
\setcounter{figure}{0}
\setcounter{table}{0}   
\setcounter{page}{1}
\makeatletter
\renewcommand{\theequation}{S\arabic{equation}}
\renewcommand{\thefigure}{S\arabic{figure}}
\renewcommand{\thetable}{S\arabic{table}}
\renewcommand{\bibnumfmt}[1]{[S#1]}
\renewcommand{\citenumfont}[1]{S#1}

\subsection{Detailed experimental configuration}

All the experiments are carried out with $\approx 10^4$ $^{87}$Rb atoms trapped in a three-beam-crossed optical dipole trap with 1003-nm wavelength. Two of the beams form a $\approx \pm11^{\circ}$ with respect to the $x$-axis (along the probe direction), while a third elliptical shaped beam travels in the $y$-axis, with all beams in the same ($x$-$y$) plane. The relative powers of the dipole beams are adjusted so that the RMS dimensions of the trapped atomic cloud are $\sigma_r=20$~$\mu$m in the radial direction and $\sigma_{x}=27$~$\mu$m.

The initial trapping and cooling take place in a magneto-optical trap (MOT). For most experiments, we load for 250 ms; if we need to adjust the atomic medium optical density (OD), we change the loading time, ranging from 50 ms to 1500 ms (with OD up to $\approx $16). Afterward, we perform a compressed-MOT stage by ramping-up the magnetic field gradient, while at the same time slowly ramping-up the dipole trap power. 

We  further cool the atoms to $\approx 10$ $\mu$K using a gray molasses~\cite{srosi2018}.

Next, we optically pump the atoms into the $\ket{5\text{S}_{1/2},F=2,m_F=2}$ state, using $\sigma_+$ polarized light blue-detuned from the $F=2$ to $F'=2$, D1 transition. We then couple the ground and Rydberg state with a two-photon transition. A 780-nm weak-probe field addresses the transition from the ground state, $\ket{5\text{S}_{1/2},F=2,m_F=2}$  to the intermediate state, $\ket{5\text{P}_{3/2},F=3,m_F=3}$; a strong-control field addresses the transition from the intermediate state to the Rydberg state, $\ket{139\text{S}_{1/2},J=1/2,m_J=1/2}$ with a wavelength of 479 nm.

Both the probe and control lasers are frequency stabilized via an ultra-low expansion (ULE) cavity with a linewidth $<10$ kHz. We use probe light that has been transmitted and filtered by the ULE cavity to reduce phase noise \cite{sleseluc2018}, .

There are eight electrodes in vacuum that allow for control of local electric fields. With this configuration, we cancel DC-Stark shifts to tens of kHz level in all three directions, shifts that would otherwise tune the Rydberg state out of resonance due to the large polarizability of the  $139S$ state, $\alpha_{139S}\approx61$ GHz/(V/cm)$^2$ \cite{sARC}.

The axial RMS of the atomic cloud, $\sigma_x\approx27$~$\mu$m is smaller than the blockade radius, $r_b\approx60$~$\mu$m to suppress the creation of multiple Rydberg atoms. Additionally, we focus the probe beam down to a $1/e^2$ waist of $w_p\approx3.3$ $\mu$m to ensure the system is effectively uni-dimensional ($w_p\leq r_b$). The control beam is counter-propagating to the probe and focused to a beam waist of $w_c\approx19$ $\mu$m. The larger beam waist provides an approximately uniform control field across the probe area.
After exiting the chamber, the probe light passes through a polarization beam splitter (PBS), and a set of bandpass filters centered at 780-nm, a narrow 1-nm bandwidth filter (Alluxa 780-1 OD6\footnote{The identification of commercial products in this paper does not imply recommendation or endorsement by the National Institute of Standards and Technology or the Army Research Laboratory, nor does it imply that the
items identified are necessarily the best available for the purpose.}), and a broader 12.5-nm bandwidth filter (Semrock LL01-780-12.5), before being coupled into a single-mode polarization-maintaining fiber (PMF). Then, the light is sent to a Hong-Ou-Mandel (HOM) interferometer, which has another set of broad filters in front of the single-mode fibers (SMF) that send the light to the single-photon avalanche detectors (SPAD) (Excelitas SPCM-780-13).  

We write a spin wave by pulsing the probe and the control field for $\approx370$~ns. The peak Rabi frequencies are $\Omega_p\approx 2\pi\times 1$ MHz and $\Omega_c\approx 2\pi\times 7$ MHz, respectively. Both fields are detuned from the intermediate state by $\Delta_p\approx2\pi\times 50$ MHz, with the two-photon transition close to resonance. Due to the collective nature from the blockaded excitation \cite{ssaffman2002}, there is a $\sqrt{N}\approx20$ enhancement to the two-photon Rabi frequency, $\sqrt{N}\Omega_{\text{2-photon}}=\sqrt{N} \Omega_p\Omega_c/(2\Delta_p)$, inferred from the $\pi$-time. This enhancement corresponds to an OD$\approx13$ given the blockaded volume.

After writing, we turn off the addressing lasers and hold (store) the spin wave in the medium for $\approx350$ ns; this is the minimum time required to switch the control acousto-optic modulator (AOM) frequency. We turn on the control field blue-detuned from the intermediate state by $\Delta_c\approx2\pi\times7$ MHz to map the spin wave into a single photon. We use an AOM before the PMF as a hardware gate to avoid saturating the SPADs from the initial write pulse.

We measure the optical losses along the path of the probe light to characterize the generation efficiency in Table~\ref{tab:transmission}.

\begin{table}[!ht]
    \centering
    \begin{tabular}{|c|c|}
        \hline
         Element &  Efficiency\\
         \hline 
         Optics transmission & 0.75(2)  \\
         \hline 
         AOM diffraction & 0.79(2)  \\
         \hline 
         PMF coupling  & 0.75(2)  \\
         \hline
         HOM-interferometer & 0.38(1) \\
         \hline
         SPAD & 0.67(1) \\
         \hline
    \end{tabular}
    \caption{List of the efficiencies along the probe path.}
    \label{tab:transmission}
\end{table}
 
The propagation efficiency includes all the optical elements, such as filters, dichroics, mirrors, polarizing beam splitters, mirrors, and lenses. With realistic improvements on higher transmission coatings and using an electro-optical modulator instead of an AOM, we could get an efficiency up to 0.65 after the PMF, from the current 0.44. 

\subsection{Background subtraction}

For all our single-photon measurements, we use two SPADs, with average background rates of $\approx80$~s$^{-1}$, and,  $\approx100$~s$^{-1}$. This count rate is due to detector dark counts and leakage of ambient light. 

Since the photons arrive at the detectors at a known time, we apply a gate corresponding to a $1.4$~$\mu$s time window, which contains more than $99.9\%$ of the pulse. We implement this in software to extract the background-photon and background-background coincidence rates from counts outside this window. With this information, we can determine the temporal profile of the accidental coincidences, which we subtract from the data. 
The probability of a background coincidence, $c_{\text{back}}$, is the sum of the products of single event rates:
\begin{equation}
c_{\text{back}}(t_1, t_2)= P_1(t_1)B_2(t_2)+B_1(t_1)P_2(t_2)+B_1(t_1)B_2(t_2),
\end{equation}

\begin{figure}[b]
\centering\includegraphics[width=0.7\textwidth]{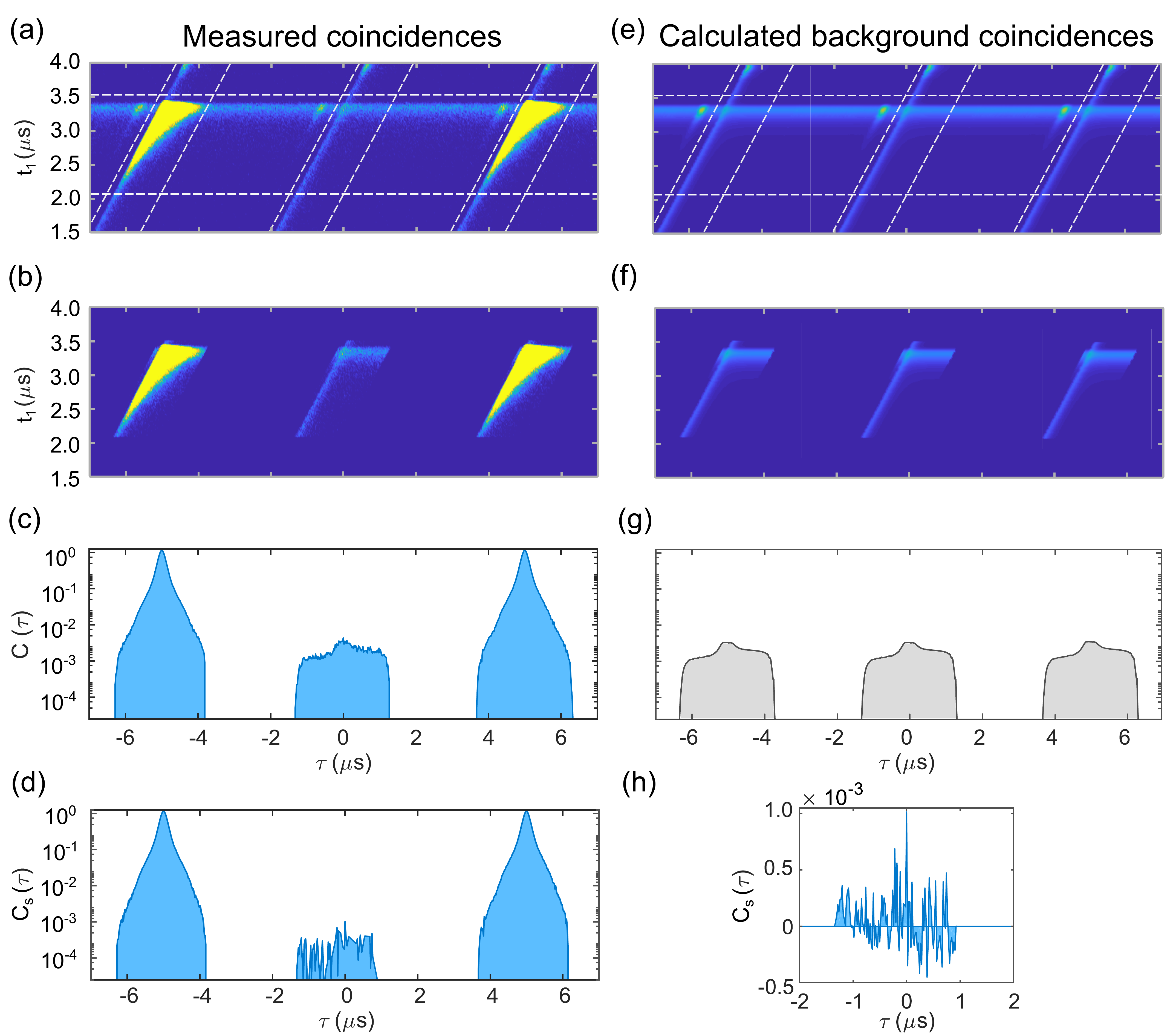}
\caption{Reconstruction of coincidences given the background and photon rate measured at each detector. (a) Raw data coincidences as a function of absolute time $t_1$, for SPAD 1 and $\tau$, the relative time between both SPADs. White-dashed lines indicate the position of the gating window for each repetition cycle. (b) Data with gate applied. (c) Total coincidence rate after applying the gate as a function of $\tau$. (d) Photon-photon coincidence rate after subtracting the background from the data. (e) Calculated background coincidence as a function of $t_1$ and $\tau$, based on the measured single-event rates $P_i(t)$ and $B_i$. (f) Calculated background coincidences after the gate. (g) Background coincidence rate as a function of $\tau$. (h) Zoom around $\tau=0$ of background-subtracted data in linear scale.} 
\label{fig:exp_coinc}
\end{figure}

where $t_1$ and $t_2$ are absolute times relative to some clock, for SPAD 1 and 2 respectively. $P_i(t_i)$, is the probability per unit time of a photon detection event at detector $i$, and $B_i(t_i)$ is the probability per unit time of a background. Changing to the relative time coordinate, $\tau=t_2-t_1$, the background coincidence probability is, 

\begin{equation}
    c_{\text{back}}(t_1, \tau)= P_1(t_1)B_2(\tau+t_1)+B_1(t_1)P_2(\tau+t_1)+B_1(t_1)B_2(\tau+t_1).
\end{equation}

We integrate $t_1$ over a time window $t_{\text{end}}-t_{\text{start}}$ to obtain the total background coincidence rate as a function of the relative time, $\tau$:
\begin{equation}
    C_{\text{back}}(\tau)=\int_{t_{\text{start}}}^{t_{\text{end}}}dt_1 \left[ P_1(t_1)B_2(\tau+t_1)+B_1(t_1)P_2(\tau+t_1)+B_1(t_1)B_2(\tau+t_1)\right],
\end{equation}
where, $t_{\text{start}}$, is synchronized to the photon arrival. With the gate, the background and pulse probability have a time dependence
\begin{equation*}
  B_1(t_1), P_1(t_1) =
    \begin{cases}
      B_1, P_1(t_1) & \text{for $t_{\text{start}}\leq t_1\leq t_{\text{end}}$}\\
      0 & \text{otherwise}
    \end{cases}       
\end{equation*}

\begin{equation*}
  B_2(\tau+t_1), P_2(\tau+t_1) =
    \begin{cases}
      B_2, P_2(\tau+t_1) & \text{for $t_1-t_{\text{end}}\leq \tau\leq t_1-t_{\text{start}}$}\\
      0 & \text{otherwise}
    \end{cases}       
\end{equation*}
With the independently measured single event rates $P_i(t)$ and $B_i$, we calculate $C_{\text{back}}(\tau)$.

This process is shown graphically in Figure~\ref{fig:exp_coinc}, where $C_{\text{back}}$ are the total coincidences rate from photon-background and background-background around $\tau=0$. Finally Figure~\ref{fig:exp_coinc}(h) shows the background subtracted coincidences rate, $C_s(\tau)$, within the gate window.

\subsection{HOM visibility discussion}

If two single photons are incident simultaneously on separate ports $a_{1}$ and $a_{2}$ of a perfect 50:50 beamsplitter (BS) the initial state $\ket{1_1, 1_2}$, becomes:
\begin{equation}
    \ket{1_1, 1_2} \rightarrow \frac{1}{\sqrt{2}} (\ket{2_3, 0_4}+\ket{0_3, 2_4})
\end{equation}
where $a_3$, $a_4$ are the output ports and we assumed that the input photons are in pure states and indistinguishable from each other. In this case, the probability of a coincidence detection is zero and the HOM visibility is one. In practice, the following factors reduce the visibility from its maximum value \cite{sStevens2013}:

\begin{itemize}
     \item one or both photons are not in a pure state, 
    \item there is more than one photon at either BS input port,
    \item an imperfect 50:50 BS. 
\end{itemize}

\noindent We will focus on the effect of the last two conditions: multi-photon events and imperfect BS.

Following the discussion from \cite{suppu2016}, we define the scattering matrix, $S$ for a general BS as,

\begin{equation}
  S=\begin{pmatrix}
    t_1 & r_2 e^{i\phi_2}\\ r_1 e^{i\phi_1} & t_2
    \end{pmatrix},
\end{equation}

\noindent where $r_1$ $(r_2)$, $t_1$ $(t_2)$, are the reflection and transmission amplitudes with a relative phase $\phi_1$ $(\phi_2)$ for port 1 (2).  

Then the input-output relations of the BS, ignoring any frequency dependence:

\begin{equation}
    \begin{pmatrix}
    \hat{a}_3\\ \hat{a}_4
    \end{pmatrix}
    =
  \begin{pmatrix}
    t_1 & r_2 e^{i\phi_2}\\ r_1 e^{i\phi_1} & t_2
    \end{pmatrix}
    \begin{pmatrix}
    \hat{a}_1\\ \hat{a}_2
    \end{pmatrix},
\end{equation}

\noindent where $\hat{a}_i$ are the photon ladder operator for the input and output ports. Generally, the scattering matrix, $S$, is not unitary.

For a lossy BS, where the output fields total energy is lower than the input fields energy, the following inequality holds:
\begin{equation}
    \sqrt{t_1^2 r_2^2 +r_1^2t_2^2+2t_1r_1t_2r_2 \cos{\alpha}} \leq \sqrt{(1-t_1^2-r_1^2)(1-t_2^2-r_2^2)},
\end{equation}
where $\alpha=\phi_1+\phi_2$, affects the maximum value that the visibility can attain. The phase, $\alpha$, is constrained by energy conservation, and we assume $\alpha=\pi$.

The number operator for the input ports 1 and 2 (output 3 and 4) is $\hat{n}_i=\hat{a}_i^{\dagger}\hat{a}_i$. Assuming that the probability of states with more than two photons is negligible, the coincidence probability, $P(1_3, 1_4)$,

\begin{equation}
   \begin{split}
    P(1_3, 1_4) & = \expval{ \hat{n}_3 \hat{n}_4}\\
    & = \expval{t_1^2 r_1^2 \hat{n}_1^2 + t_2^2r_2^2 \hat{n}_2^2+(t_1^2t_2^2+r_1^2r_2^2-2t_1r_1t_2r_2)\hat{n}_1\hat{n}_2} \\
    & = (t_1^2 r_1^2 +  t_2^2r_2^2) 2 P_2 + (t_1^2t_2^2+r_1^2r_2^2-2 c t_1r_1t_2r_2)P_1^2.
\end{split}
\end{equation}
Here $P_1$ is the probability of a single photon, $P_2$ is the probability of two photons at one input port, and $c$ is the mode overlap of the two incident photons. Following the assumption that the probability of more than two-photon states is negligible, we can rewrite $P_2$ as a function of the correlation function $g^{(2)}(0)$ and $P_1$, as $P_2\approx g^{(2)}(0) P_1^2 /2$. The coincidence probability:
\begin{equation}
   P(1_3, 1_4) \approx \left[t_1^2t_2^2+r_1^2r_2^2+(t_1^2 r_1^2 +  t_2^2r_2^2) g^{(2)} -2 c t_1r_1t_2r_2 \right] P_1^2.
\end{equation}

For the more general case, where the BS coefficients are not the same for orthogonal polarizations, $H$, and $V$

\begin{equation}
   \begin{split}
  \mathcal{V} &= \frac{\eval{P(1_3, 1_4)_{HV}}_{c=0}-\eval{P(1_3, 1_4)_{HH}}_{c=c}}{\eval{P(1_3, 1_4)_{HV}}_{c=0}}\\ 
    & = \frac{t_{1_V}^2t_{2_H}^2+r_{1_V}^2r_{2_H}^2-t_{1_H}^2t_{2_H}^2-r_{1_H}^2r_{2_H}^2+(t_{1_V}^2 r_{1_V}^2 -t_{1_H}^2 r_{1_H}^2) g^{(2)} +2 c t_{1_H}r_{1_H}t_{2_H}r_{2_H}}{t_{1_V}^2t_{2_H}^2+r_{1_V}^2r_{2_H}^2+(t_{1_V}^2 r_{1_V}^2 +  t_{2_H}^2r_{2_H}^2)g^{(2)}},
\end{split}
\end{equation}
where we assume that in the case of $P(1_3, 1_4)_{HV}$, the photon at port 1 has $H$-polarization and the photon at port 2 has $V$-polarization, similarly for $P(1_3, 1_4)_{HH}$, both incoming photons have $H-$polarization.

In the particular case of a BS with 
symmetric ports, $t_1^2=t_2^2=T$ and, $r_1^2=r_2^2=R$, the visibility reduces to:

\begin{equation}
   \mathcal{V}=\frac{ 2 c}{T/R+R/T+2 g^{(2)}}.
\end{equation}
If $T=R=1/2$ and $g^{(2)}(0)=0$, then the visibility is equal to the incoming photons overlap, $c$.

In the following table, we show the measured transmission and reflection coefficients of the BS used in the HOM characterization, for both $H$- and $V$-polarization:

\begin{table}[!ht]
    \centering
    \begin{tabular}{|c|c|c|}
        \hline
         Port/Polarization &  $T$ & $R$ \\
         \hline 
         Port 1 $H$ & 0.502(5) & 0.421(3) \\
         \hline 
         Port 1 $V$ & 0.484(5) & 0.428(3) \\
         \hline 
         Port 2 $H$ & 0.511(9) & 0.426(5) \\
         \hline
    \end{tabular}
    \caption{Transmission and reflection coefficients for the BS used in the HOM interferometer.}
    \label{tab:BS_coeff}
\end{table}

We measured a background-subtracted visibility to be $\mathcal{V}=0.966(6)$, and using equation (13) to take into account the imperfect BS, we find a mode overlap of 0.982(7).

\subsection{Contaminants}

We use a simple model to characterize the effects of the contaminants on the photon generation, where there is a probability that a stored spin wave is converted to a contaminant. Once a contaminant is present in the medium, it disables the writing and storing of a spin wave until the contaminant decays, with a time constant $\tau_c$. If $P_c$ is the probability of creating a contaminant on a given pulse, then the probability, $P_n$, of a contaminant being present at pulse $n$ depends on whether one was created in one of the previous pulses and remained to the $n$-th pulse
\begin{equation}
    P_n=P_{n-1}e^{-t_p/\tau_c}+(1-P_{n-1})P_c,
\end{equation}
where $t_p$ is the pulse spacing. If we set the initial condition to be $P_1=P_c$, and use the identity,  $(1-x)\sum_{j=0}^{n-1}x^j=1-x^n$, we get the expression:

\begin{equation}
    P_n=P_c \frac{1-(e^{- t_p/\tau_c}-P_c)^n  }{1-e^{-t_p/\tau_c}+P_c}.
\end{equation}
Then, the probability of successfully generating a photon, $P_g (n)$ is
\begin{equation}
    P_g(n)=P_{max}(1-P_n)=P_{max}\left( 1 - P_c \frac{1-(e^{- t_p/\tau_c}-P_c)^n  }{1-e^{-t_p/\tau_c}+P_c}\right),
\end{equation}
where $P_{max}$ is the maximum probability of generating a photon. For $n\rightarrow\infty$, the steady state probability $P_s$,
\begin{equation}
    P_s\approx P_{max}\left( 1 - \frac{P_c}{1-e^{-t_p/\tau_c}+P_c}\right).
\end{equation}

We also model how the correlation function, $g^{(2)}(m\ t_p)$ for integer $m\neq0$, is modified due to contaminants:
\begin{equation}
\begin{split}
g^{(2)}(|m| t_p)&=\frac{ \langle P_s P_g(m)\rangle }{ \langle P_s^2 \rangle }   \\
 &=1+P_c\frac{(e^{-t_p/\tau_c}-P_c)^n}{1-e^{-t_p/\tau_c}},
\end{split}
\label{eq:g2}
\end{equation}
this manifests as a bunching feature around $\tau=0$.

\subsection{Theoretical model}

\subsubsection{Write and storage efficiency}

\begin{figure}[ht]
\centering\includegraphics[width=0.2\textwidth]{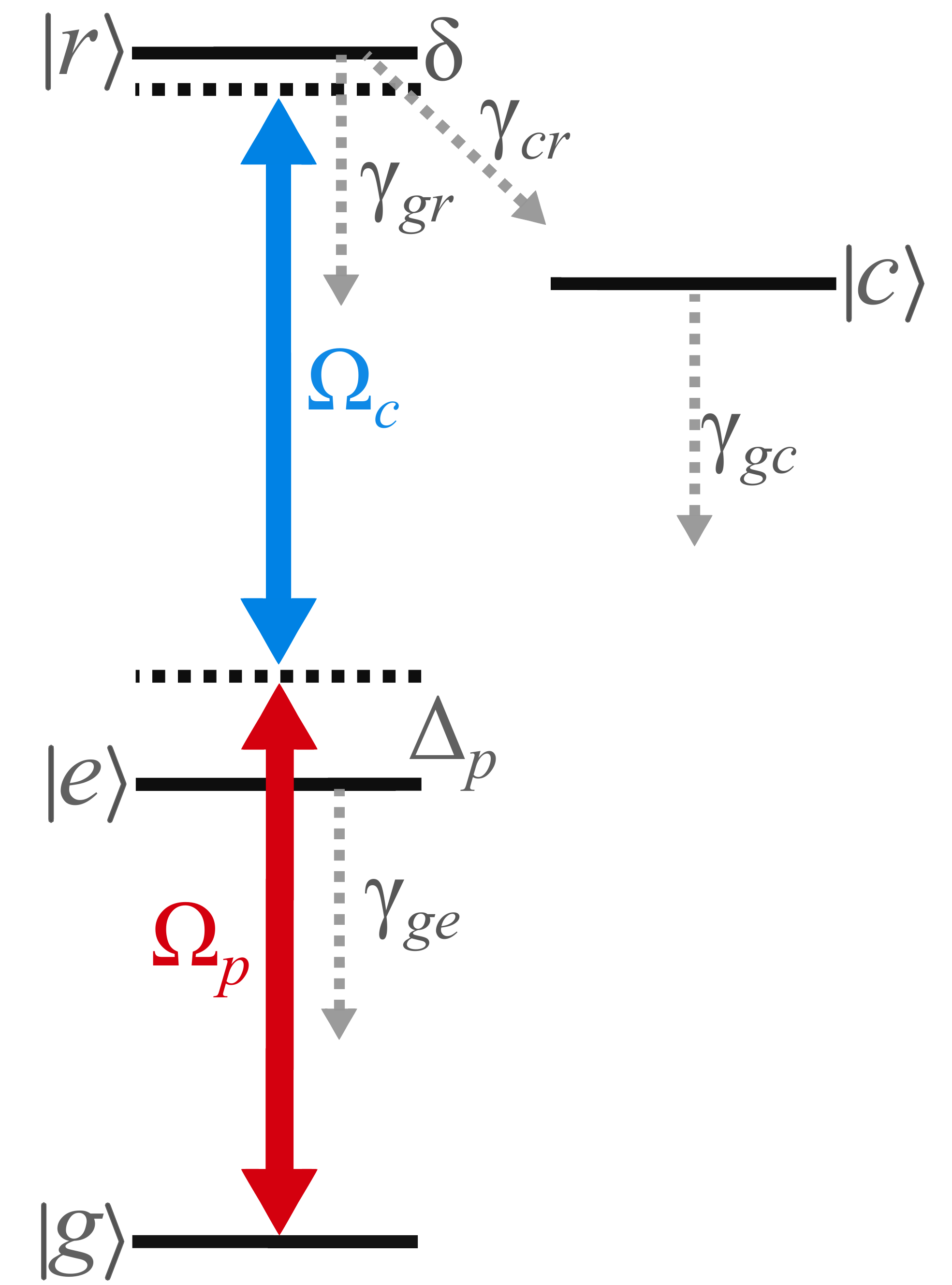}
\caption{Atomic levels showing the driving fields and decay rates used in theoretical model. Here we show the ground-state, as $\ket{g}=\ket{5S_{1/2}}$, the intermediate state $\ket{e}=\ket{5P_{3/2}}$, Rydberg state $\ket{r}=\ket{139S_{1/2}}$, and the contaminants states as $\ket{c}$.}
\label{fig:levels}
\end{figure}

We model the spin-wave as a super-atom with $N$-atoms being collectively driven into a single excitation to the Rydberg state, for the writing and storage time. The energy levels and decay rates of the super-atom are shown in Figure \ref{fig:levels}.

We simulated the writing stage as driving the super-atom from the ground to the Rydberg state, with $\sqrt{N}$-enhanced Rabi frequency. During the writing time, $t_{w}$, the Rabi frequencies, $\Omega_p\approx2\pi\times1.0(2)$~MHz and $\Omega_c\approx2\pi\times6.8(3)$~MHz are kept constant. For the storage time, $t_s$, these driving frequencies are set to zero. 

The Hamiltonian describing the the system depicted in Fig~\ref{fig:levels} in the rotating wave approximation is given by: 

\begin{eqnarray}
H(t) &=&\frac{\hbar}{2}
\!
	\left(\begin{array}{cccc}
	0 & \sqrt{N} \Omega_p(t) & 0 & 0 \\
	\sqrt{N} \Omega_p(t)  & -2 \Delta_p & \Omega_c(t) & 0 \\
	0 & \Omega_c(t) & -2\delta & 0 \\
	0 & 0 & 0&0
	\end{array}
	\right)
\label{eq:H},
\end{eqnarray}
in the basis of $\ket{g}$, $\ket{e}$, $\ket{r}$, $\ket{c}$, for the ground, intermediate, Rydberg and contaminant state, respectively.

Using the Python package QuTip \cite{squtip}, we calculated the non-unitary dynamics of this first stage using the master equation for the four level density matrix $\rho$:

\begin{equation}
    \dot{\rho}=-\frac{i}{\hbar} [H, \rho]-\sum_n \frac{1}{2} \{\rho, C_n^{\dagger}C_n\}+C_n\rho C_n^{\dagger},
    \label{eq:Lind}
\end{equation}
where $C_1=\sqrt{\gamma_{ge}}\ket{g}\bra{e}$, $C_2=\sqrt{\gamma_{gr}}\ket{g}\bra{r}$, $C_3=\sqrt{\gamma_{cr}}\ket{c}\bra{r}$, and $C_4=\sqrt{\gamma_{gc}}\ket{g}\bra{c}$ are the jump operators. 

Given the decay rates of the different states: $\gamma_{ge}\approx2\pi\times6.9(6)$ MHz, $\gamma_{gr}\approx2\pi\times88(6)$ kHz, $\gamma_{cr}\approx2\pi\times5(1)$ kHz, and $\gamma_{gc}\approx2\pi\times2.5(3)$ kHz, we calculate that the writing and storage efficiency are $\eta_w=0.82(1)$, $\eta_s=0.82(1)$, respectively.  

\subsubsection{Retrieval efficiency}
We follow the derivations in  Ref.~\cite{salexey2007} to compute the retrieval efficiency.
In the rescaled unit-less coordinates, $\tilde{z}=0$ and $\tilde{z}=1$ represent the front and the end of the atomic cloud, respectively. Suppose all atoms are in the $|r\rangle$ state in the beginning of the retrieval stage at time $\tilde{t}=0$, the shape of the spin wave is given by $S(\tilde{z},\tilde{t}=0) = 1$ for $\tilde{z}\in [0, 1]$ and $S(\tilde{z},\tilde{t}=0) = 0$ for $\tilde{z}$ elsewhere. 
The retrieval efficiency can be expressed in terms of the photon field $\mathcal{E}(\tilde{z},\tilde{t})$ emitted by the stored spin wave at the end of the atomic cloud:
\begin{align}
    \eta_r&=\int_0^{\infty}d\tilde{t}|\mathcal{E}(\tilde{z}=1, \tilde{t})|^2.\label{eqetar}
\end{align}
$\mathcal{E}(1,\tilde{t})$ can be calculated as:
\begin{equation}
\mathcal{E}(1,\tilde{t})=-\sqrt{d}\tilde{\Omega}(\tilde{t})\exp(-\tilde{\gamma}_s\tilde{t})
\int_{0}^{1}d\tilde{z}\frac{1}{1+i\tilde{\Delta}}e^{-(h(0,\tilde{t})+d\tilde{z})/(1+i\tilde{\Delta})} I_0\left(2\frac{\sqrt{h(0,\tilde{t})d\tilde{z}}}{1+i\tilde{\Delta}}\right)S(1-\tilde{z}),
\end{equation}
where we define dimensionless parameters $d=$OD/2,  $\tilde{\gamma}_s=(\gamma_{gr}+\gamma_{cr})/\gamma_{ge}$, $\tilde{\Delta}=2\Delta_p/\gamma_{ge}$, $\tilde{\Omega}(t)=\Omega_c(t)/\gamma_{ge}$. $h(\tilde{t},\tilde{t}')=\int^{\tilde{t}'}_{\tilde{t}}|\tilde{\Omega}(\tilde{t}'')|^2d\tilde{t}''$ and $I_0$ is the $0th$-order modified Bessel function of the first kind.
When the control field $\Omega_c$ is constant in time, we define the dimensionless parameter $x_s=2\tilde{\gamma_{s}}/|\tilde{\Omega}_c|^2$ which characterizes the strength of the decay rate compared to the control field.  \eqref{eqetar} can be evaluated as
\begin{equation}
   \eta_r =\int_{0}^{1}d\bar{z}\int_{0}^{1}d\bar{z}'K_r S(1-\bar{z}) S^*(1-\bar{z}'),
\label{eq:integral}
\end{equation}
where $K_r$ is given by
\begin{equation}
     K_r=\frac{d f(x_s)}{2}\exp\left[ -\frac{d f(x_s)}{2}\left( (1+x_s(1-i\tilde{\Delta}))\bar{z}+(1+x_s(1+i\tilde{\Delta}))d\bar{z}'\right)\right]I_0(d\sqrt{\bar{z}\bar{z}'}f(x_s)),\label{eq:kr}
\end{equation}
and $f(x_s)=\frac{2}{2+x_s(1+\tilde{\Delta}^2)}$.
 
Evaluating the integral in Eq.\ \eqref{eq:integral} numerically, we obtain the retrieval efficiency $\eta_r=0.63(2)$. With these results, we estimate that the photon generation probability at the end of the cloud is $P_{\text{th}}=0.42(3)$.
 
 \subsubsection{Possible improvements}
With conservative feasible experimental improvements, such as implementing a ground-state blue-detuned optical dipole trap, as well as increasing the following parameters: $\Omega_c=2\pi\times10$~MHz, $\Delta_p=2\pi\times100$~MHz and OD=20, while decreasing the spin wave dephasing by a factor of two, we estimate that we could increase our probabilities up to $\eta_w\eta_s=0.86$ and $\eta_r=0.72$, while maintaining a relatively low contaminant probability, $P_c\approx3\times10^{-2}$.
 
From the theoretical model, the main limiting factor is the retrieval process; in principle, the retrieval efficiency increases with higher OD; however, the contaminant production also grows with OD. A Rydberg ensemble with low OD coupled to a cavity could further increase light-matter interactions and therefore increase the overall photon production probability, making it a promising platform for scalable quantum information applications. 

\subsection{Single-photon sources}

In Tables \ref{tab:SS_source} and  \ref{tab:atom_source}, there is detailed information about the properties of a representative sample of single-photon sources plotted in Fig.~5. in the main text. The notation, $R$, repetition rate, $P$ is the probability of coupling a single-photon into a single-mode fiber, $V$, is the indistinguishability, $\eta$ is the single-mode probability, $\mathcal{R}$ is the brightness, and $\mathcal{F}$ is the fidelity.

\begin{table*}[h]
    \centering
    \begin{tabular}{|m{0.07\textwidth}|m{0.05\textwidth}|m{0.06\textwidth}|m{0.06\textwidth}|m{0.06\textwidth}|m{0.06\textwidth}|m{0.06\textwidth}|m{0.09\textwidth}|m{0.06\textwidth}|}
        \hline
        Type & Ref & $R$ (MHz) & $P$ &  $V$ & $g^{(2)}$ & $\eta$ & $\mathcal{R}$ $\times10^6 (s^{-1})$ & $\mathcal{F}$ \\
        \hline 
        SPDC & \cite{swang2016} & 76 & $\approx$0.01 & 0.91 & 0.09 & 0.009 & 0.69 & 0.910 \\
        \hline 
        MUX & \cite{sxiong2016} & 10 & $\approx$0.002 & 0.91 & $\sim$0.2 & 0.002 & 0.02 & 0.910 \\
        \hline
        MUX & \cite{skaneda2019}  & 0.5 & 0.667 & 0.91 & 0.269 & 0.561 & 0.28 &0.840 \\
        \hline
        QD & \cite{ssomaschi2016}  & 82 & $\approx$0.02 & 0.996 & 0.024 & 0.020 & 1.63 & 0.996 \\
        \hline
        QD &\cite{sloredo2016} & 80 & 0.024 & 0.7 & 0.013 & 0.017 & 1.34 &0.699 \\
        \hline
        QD & \cite{swang2017} & 76 & 0.337 & 0.93 & 0.027 & 0.312 & 23.71 &0.926 \\
        \hline
        QD & \cite{skir2017} & 76 & 0.10 & 0.94 & 0.006 & 0.094 & 7.14& 0.940 \\
       \hline
        QD & \cite{swang2019p} & 76 & 0.24 & 0.975 & 0.025 & 0.233 & 17.7& 0.972\\
        \hline
    \end{tabular}
    \caption{Table comparing the performances of solid state single-photon sources: spontaneous parametric down conversion (SPDC), multiplexed-heralded-single-photon source (MUX) and, quantum dot (QD). Values estimated from available data.}
    \label{tab:SS_source}
\end{table*}

\begin{table*}[h]
    \centering
    \begin{tabular}{|m{0.1\textwidth}|m{0.05\textwidth}|m{0.08\textwidth}|m{0.06\textwidth}|m{0.06\textwidth}|m{0.06\textwidth}|m{0.06\textwidth}|m{0.06\textwidth}|m{0.09\textwidth}|m{0.06\textwidth}|}
        \hline
        Type & Ref & Duty Cycle ($\%$) & $R$ (MHz)&  $P$ &  $V$ & $g^{(2)}$ & $\eta$ & $\mathcal{R}$ $\times10^3 (s^{-1})$ & $\mathcal{F}$ \\
        \hline 
        Yb ion & \cite{smaunz2007} & 80 & 8 & 0.003 & 0.86 & $\sim10^{-3}$ & 0.003 & 18.16 & 0.860 \\
        \hline
        Rb Atom &\cite{srosenfeld2017} & 33 & 0.052 & 0.003 & 0.9 & $\sim10^{-3}$ & 0.003 & 0.05& 0.899\\
        \hline
        Ensemble in cavity & \cite{sthompson2006} & $\approx1.8$ & 0.05 & 0.08 & 0.9 & 0.05 & 0.072 &0.06 &0.898 \\
        \hline 
        Atom in cavity & \cite{swilk2007} & $\approx2$ & 0.7 &  0.2 & 0.7 & $\sim10^{-2}$ & 0.140 &1.96& 0.699 \\
        \hline
        Atom in cavity &\cite{snisbet2011} & 0.1 & 1 & 0.21 & 0.87 & 0.02 & 0.182 &0.18& 0.868\\
        \hline
       Atom in cavity &\cite{smucke2013} & 100 & 0.01 & 0.39 & 0.64 & 0.02 & 0.249 &2.49&0.637 \\
        \hline
        Rydberg & this work & 60 & 0.013 & 0.141 & 0.982 & $\approx10^{-4}$ & 0.139 & 1.11 & 0.982\\
        \hline
        Rydberg  & future & 60 & 0.5 & 0.4 & 0.99 & $\approx10^{-4}$ & $\approx0.4$&120&0.99\\
        \hline
    \end{tabular}
    \caption{Table comparing the performances of different atomic single-photon sources. Here $\mathcal{R}$ is weighted by the duty cycle of operation. Values estimated from available data.}
    \label{tab:atom_source}
\end{table*}


\begin{thebibliography}{10}
\newcommand{\enquote}[1]{``#1''}

\bibitem{carolan2015}
J.~Carolan, C.~Harrold, C.~Sparrow, E.~Mart{\'\i}n-L{\'o}pez, N.~J. Russell,
  J.~W. Silverstone, P.~J. Shadbolt, N.~Matsuda, M.~Oguma, M.~Itoh, G.~D.
  Marshall, M.~G. Thompson, J.~C.~F. Matthews, T.~Hashimoto, J.~L.
  O{\textquoteright}Brien, and A.~Laing, \enquote{Universal linear optics,}
  {{Science}} \textbf{349}, 711--716 (2015).

\bibitem{wang2019}
H.~Wang, J.~Qin, X.~Ding, M.-C. Chen, S.~Chen, X.~You, Y.-M. He, X.~Jiang,
  L.~You, Z.~Wang, C.~Schneider, J.~J. Renema, S.~H\"ofling, C.-Y. Lu, and
  J.-W. Pan, \enquote{Boson sampling with 20 input photons and a 60-mode
  interferometer in a $1{0}^{14}$-dimensional hilbert space,}
  {{Phys. Rev. Lett.}} \textbf{123}, 250503 (2019).

\bibitem{yin2017}
J.~Yin, Y.~Cao, Y.-H. Li, S.-K. Liao, L.~Zhang, J.-G. Ren, W.-Q. Cai, W.-Y.
  Liu, B.~Li, H.~Dai \emph{et~al.}, \enquote{Satellite-based entanglement
  distribution over 1200 kilometers,} {{Science}}
  \textbf{356}, 1140--1144 (2017).

\bibitem{slussarenko2017}
S.~Slussarenko, M.~M. Weston, H.~M. Chrzanowski, L.~K. Shalm, V.~B. Verma,
  S.~W. Nam, and G.~J. Pryde, \enquote{Unconditional violation of the
  shot-noise limit in photonic quantum metrology,}
  {{Nature Photon}} \textbf{11}, 700--703 (2017).

\bibitem{wang2019efficient}
Y.~Wang, J.~Li, S.~Zhang, K.~Su, Y.~Zhou, K.~Liao, S.~Du, H.~Yan, and S.-L.
  Zhu, \enquote{Efficient quantum memory for single-photon polarization
  qubits,} {{Nature Photon}} \textbf{13}, 346--351 (2019).

\bibitem{yu2020}
Y.~Yu, F.~Ma, X.-Y. Luo, B.~Jing, P.-F. Sun, R.-Z. Fang, C.-W. Yang, H.~Liu,
  M.-Y. Zheng, X.-P. Xie \emph{et~al.}, \enquote{Entanglement of two quantum
  memories via fibres over dozens of kilometres,}
  {{Nature}} \textbf{578}, 240--245 (2020).

\bibitem{bock2018}
M.~Bock, P.~Eich, S.~Kucera, M.~Kreis, A.~Lenhard, C.~Becher, and J.~Eschner,
  \enquote{High-fidelity entanglement between a trapped ion and a telecom
  photon via quantum frequency conversion,} {{Nat Commun}}
  \textbf{9}, 1--7 (2018).

\bibitem{ballance2016}
C.~J. Ballance, T.~P. Harty, N.~M. Linke, M.~A. Sepiol, and D.~M. Lucas,
  \enquote{High-fidelity quantum logic gates using trapped-ion hyperfine
  qubits,} {{Phys. Rev. Lett.}} \textbf{117}, 060504
  (2016).

\bibitem{gross2017}
C.~Gross and I.~Bloch, \enquote{Quantum simulations with ultracold atoms in
  optical lattices,} {{Science}} \textbf{357}, 995--1001
  (2017).

\bibitem{peyronel2012}
T.~Peyronel, O.~Firstenberg, Q.-Y. Liang, S.~Hofferberth, A.~V. Gorshkov,
  T.~Pohl, M.~D. Lukin, and V.~Vuleti{\'c}, \enquote{Quantum nonlinear optics
  with single photons enabled by strongly interacting atoms,}
  {{Nature}} \textbf{488}, 57--60 (2012).

\bibitem{maxwell2013}
D.~Maxwell, D.~Szwer, D.~Paredes-Barato, H.~Busche, J.~D. Pritchard,
  A.~Gauguet, K.~J. Weatherill, M.~Jones, and C.~S. Adams, \enquote{Storage and
  control of optical photons using rydberg polaritons,}
  {{Phys. Rev. Lett.}} \textbf{110}, 103001 (2013).

\bibitem{li2016}
L.~Li and A.~Kuzmich, \enquote{Quantum memory with strong and controllable
  rydberg-level interactions,} {{Nat Commun}} \textbf{7},
  13618 (2016).

\bibitem{paris2017}
A.~Paris-Mandoki, C.~Braun, J.~Kumlin, C.~Tresp, I.~Mirgorodskiy,
  F.~Christaller, H.~P. B{\"u}chler, and S.~Hofferberth, \enquote{Free-space
  quantum electrodynamics with a single rydberg superatom,}
  {{Physical Review X}} \textbf{7}, 041010 (2017).

\bibitem{dudin2012}
Y.~Dudin and A.~Kuzmich, \enquote{Strongly interacting rydberg excitations of a
  cold atomic gas,} {{Science}} \textbf{336}, 887--889
  (2012).

\bibitem{ripka2018}
F.~Ripka, H.~K{\"u}bler, R.~L{\"o}w, and T.~Pfau, \enquote{A room-temperature
  single-photon source based on strongly interacting rydberg atoms,}
  {{Science}} \textbf{362}, 446--449 (2018).

\bibitem{gorni2014}
H.~Gorniaczyk, C.~Tresp, J.~Schmidt, H.~Fedder, and S.~Hofferberth,
  \enquote{Single-photon transistor mediated by interstate rydberg
  interactions,} {{Phys. Rev. Lett.}} \textbf{113}, 053601
  (2014).

\bibitem{tiarks2014}
D.~Tiarks, S.~Baur, K.~Schneider, S.~D{\"u}rr, and G.~Rempe,
  \enquote{Single-photon transistor using a f{\"o}rster resonance,}
  {{Phys. Rev. Lett.}} \textbf{113}, 053602 (2014).

\bibitem{gorni2016}
H.~Gorniaczyk, C.~Tresp, P.~Bienias, A.~Paris-Mandoki, W.~Li, I.~Mirgorodskiy,
  H.~B{\"u}chler, I.~Lesanovsky, and S.~Hofferberth, \enquote{Enhancement of
  rydberg-mediated single-photon nonlinearities by electrically tuned
  f{\"o}rster resonances,} {{Nat Commun}} \textbf{7},
  12480 (2016).

\bibitem{tiarks2016}
D.~Tiarks, S.~Schmidt, G.~Rempe, and S.~D{\"u}rr, \enquote{Optical $\pi$ phase
  shift created with a single-photon pulse,} {{Science
  Advances}} \textbf{2}, e1600036 (2016).

\bibitem{thompson2017}
J.~D. Thompson, T.~L. Nicholson, Q.-Y. Liang, S.~H. Cantu, A.~V. Venkatramani,
  S.~Choi, I.~A. Fedorov, D.~Viscor, T.~Pohl, M.~D. Lukin \emph{et~al.},
  \enquote{Symmetry-protected collisions between strongly interacting photons,}
  {{Nature}} \textbf{542}, 206--209 (2017).

\bibitem{tiarks2018}
D.~Tiarks, S.~Schmidt-Eberle, T.~Stolz, G.~Rempe, and S.~D{\"u}rr, \enquote{A
  photon--photon quantum gate based on rydberg interactions,}
  {{Nature Phys}} \textbf{15}, 124--126 (2018).

\bibitem{maller2015}
K.~Maller, M.~Lichtman, T.~Xia, Y.~Sun, M.~Piotrowicz, A.~Carr, L.~Isenhower,
  and M.~Saffman, \enquote{Rydberg-blockade controlled-not gate and
  entanglement in a two-dimensional array of neutral-atom qubits,}
  {{Physical Review A}} \textbf{92}, 022336 (2015).

\bibitem{zeng2017}
Y.~Zeng, P.~Xu, X.~He, Y.~Liu, M.~Liu, J.~Wang, D.~Papoular, G.~Shlyapnikov, and M.~Zhan, 
\enquote{Entangling two individual atoms of different isotopes via rydberg blockade,} {{Phys.~Rev.~Lett.}}
  \textbf{119}, 160502 (2017).

\bibitem{levine2018} 
H.~Levine, A.~Keesling, A.~Omran, H.~Bernien, S.~Schwartz, A.~S.~Zibrov, M.~Endres, M.~Greiner, V.~Vuleti{\'c}, and M.~D.~Lukin, 
\enquote{High-fidelity control and entanglement of rydberg-atom qubits,} {{Phys.~Rev.~Lett.}} \textbf{121}, 123603 (2018).

\bibitem{us}
A.~N. Craddock, J.~Hannegan, D.~P. Ornelas-Huerta, J.~D. Siverns, A.~J.
  Hachtel, E.~A. Goldschmidt, J.~V. Porto, Q.~Quraishi, and S.~L. Rolston,
  \enquote{Quantum interference between photons from an atomic ensemble and a
  remote atomic ion,} {{Phys. Rev. Lett.}} \textbf{123},
  213601 (2019).

\bibitem{schauss2012}
P.~Schau{\ss}, M.~Cheneau, M.~Endres, T.~Fukuhara, S.~Hild, A.~Omran, T.~Pohl,
  C.~Gross, S.~Kuhr, and I.~Bloch, \enquote{Observation of spatially ordered
  structures in a two-dimensional rydberg gas,} {{Nature}}
  \textbf{491}, 87--91 (2012).

\bibitem{zeiher2017}
J.~Zeiher, J.-y. Choi, A.~Rubio-Abadal, T.~Pohl, R.~van Bijnen, I.~Bloch, and
  C.~Gross, \enquote{Coherent many-body spin dynamics in a long-range
  interacting ising chain,} {{Physical Review X}}
  \textbf{7}, 041063 (2017).

\bibitem{lienhard2018}
V.~Lienhard, S.~De~L{\'e}s{\'e}leuc, D.~Barredo, T.~Lahaye, A.~Browaeys,
  M.~Schuler, L.-P. Henry, and A.~M. L{\"a}uchli, \enquote{Observing the
  space-and time-dependent growth of correlations in dynamically tuned
  synthetic ising models with antiferromagnetic interactions,}
  {{Physical Review X}} \textbf{8}, 021070 (2018).

\bibitem{kim2018}
H.~Kim, Y.~Park, K.~Kim, H.-S. Sim, and J.~Ahn, \enquote{Detailed balance of
  thermalization dynamics in rydberg-atom quantum simulators,}
  {{Phys. Rev. Lett.}} \textbf{120}, 180502 (2018).

\bibitem{saffman2002}
M.~Saffman and T.~G. Walker, 
\enquote{Creating single-atom and single-photon sources from entangled atomic ensembles,} {{Phys. Rev. A}} \textbf{66}, 065403 (2002).

\bibitem{lukin2001}
M.~D. Lukin, M.~Fleischhauer, R.~Cote, L.~M. Duan, D.~Jaksch, J.~I. Cirac, and  P.~Zoller, \enquote{Dipole blockade and quantum information processing in mesoscopic atomic ensembles,} {{Phys. Rev. Lett.}} \textbf{87}, 037901 (2001).
  
\bibitem{sangouard2011} 
N.~Sangouard, C.~Simon, H.~de~Riedmatten, and N.~Gisin,
\enquote{Quantum repeaters based on atomic ensembles and linear optics,} {{Rev.~Mod.~Phys.}}~\textbf{83}, 33--80 (2011).

\bibitem{Elizabeth2016} 
E.~A. Goldschmidt, T.~Boulier, R.~C. Brown, S.~B. Koller, J.~T. Young, A.~V. Gorshkov, S.~L. Rolston, and J.~V. Porto, \enquote{Anomalous broadening in driven dissipative rydberg systems,} {{Phys. Rev. Lett.}} \textbf{116}, 113001 (2016).

\bibitem{eisaman2011}
M.~D. Eisaman, J.~Fan, A.~Migdall, and S.~V. Polyakov, \enquote{Invited review article: Single-photon sources and detectors,} {{Review of Scientific Instruments}} \textbf{82}, 071101 (2011).


\bibitem{ARC}
N.~{\v{S}}ibali{\'c}, J.~D. Pritchard, C.~S. Adams, and K.~J. Weatherill,
  \enquote{Arc: An open-source library for calculating properties of alkali
  rydberg atoms,} {{Computer Physics Communications}}
  \textbf{220}, 319--331 (2017).

\bibitem{dudin2012coll}
Y.~Dudin, L.~Li, F.~Bariani, and A.~Kuzmich, \enquote{Observation of coherent
  many-body rabi oscillations,} {{Nature Physics}}
  \textbf{8}, 790--794 (2012).

\bibitem{SM}Supplemental Material contains the details on experimental configuration, background subtraction, HOM visibility reduction discussion, contaminant states creation, theory model to estimate writing and retrieval efficiencies, possible improvements, and information about single-photon sources plotted in Fig.~\ref{fig:comparison}.

\bibitem{alexey2007}
A.~V. Gorshkov, A.~Andr\'e, M.~D. Lukin, and A.~S. S\o{}rensen, \enquote{Photon
  storage in $\ensuremath{\Lambda}$-type optically dense atomic media. ii.
  free-space model,} {{Phys. Rev. A}} \textbf{76}, 033805
  (2007).

\bibitem{DeSalvo2016}
B.~J. DeSalvo, J.~A. Aman, C.~Gaul, T.~Pohl, S.~Yoshida, J.~Burgd\"orfer,
  K.~R.~A. Hazzard, F.~B. Dunning, and T.~C. Killian, \enquote{Rydberg-blockade
  effects in autler-townes spectra of ultracold strontium,}
  {{Phys. Rev. A}} \textbf{93}, 022709 (2016).

\bibitem{Radiation2017}
D.~P. Sadler, E.~M. Bridge, D.~Boddy, A.~D. Bounds, N.~C. Keegan, G.~Lochead,
  M.~P.~A. Jones, and B.~Olmos, \enquote{Radiation trapping in a dense cold
  rydberg gas,} {{Phys. Rev. A}} \textbf{95}, 013839
  (2017).

\bibitem{Aman2016}
J.~A. Aman, B.~J. DeSalvo, F.~B. Dunning, T.~C. Killian, S.~Yoshida, and
  J.~Burgd\"orfer, \enquote{Trap losses induced by near-resonant rydberg
  dressing of cold atomic gases,} {{Phys. Rev. A}}
  \textbf{93}, 043425 (2016).

\bibitem{Chem2016}
M.~Schlagm\"uller, T.~C. Liebisch, F.~Engel, K.~S. Kleinbach, F.~B\"ottcher,
  U.~Hermann, K.~M. Westphal, A.~Gaj, R.~L\"ow, S.~Hofferberth, T.~Pfau,
  J.~P\'erez-R\'{\i}os, and C.~H. Greene, \enquote{Ultracold chemical reactions
  of a single rydberg atom in a dense gas,} {{Phys. Rev.
  X}} \textbf{6}, 031020 (2016).

\bibitem{boulier2017}
T.~Boulier, E.~Magnan, C.~Bracamontes, J.~Maslek, E.~A. Goldschmidt, J.~T.
  Young, A.~V. Gorshkov, S.~L. Rolston, and J.~V. Porto, \enquote{Spontaneous
  avalanche dephasing in large rydberg ensembles,} {{Phys.
  Rev. A}} \textbf{96}, 053409 (2017).

\bibitem{bienias2018} 
P.~{Bienias}, J.~{Douglas}, A.~{Paris-Mandoki}, P.~{Titum}, I.~{Mirgorodskiy}, C.~{Tresp}, E.~{Zeuthen}, M.~J. {Gullans}, M.~{Manzoni}, S.~{Hofferberth}, D.~{Chang}, and A.~V. {Gorshkov}, \enquote{{Photon propagation through dissipative Rydberg media at large input rates},} {{arXiv e-prints}} arXiv:1807.07586 (2018).

\bibitem{wang2016}
X.-L. Wang, L.-K. Chen, W.~Li, H.-L. Huang, C.~Liu, C.~Chen, Y.-H. Luo, Z.-E. Su, D.~Wu, Z.-D. Li, H.~Lu, Y.~Hu, X.~Jiang, C.-Z. Peng, L.~Li, N.-L. Liu, Y.-A. Chen, C.-Y. Lu, and J.-W. Pan, \enquote{Experimental ten-photon entanglement,} {{Phys. Rev. Lett.}} \textbf{117}, 210502 (2016).

\bibitem{xiong2016}
C.~Xiong, X.~Zhang, Z.~Liu, M.~J. Collins, A.~Mahendra, L.~Helt, M.~J. Steel,
  D.-Y. Choi, C.~Chae, P.~Leong \emph{et~al.}, \enquote{Active temporal
  multiplexing of indistinguishable heralded single photons,}
  {{Nat Commun}} \textbf{7}, 10853 (2016).

\bibitem{kaneda2019}
F.~Kaneda and P.~G. Kwiat, \enquote{High-efficiency single-photon generation
  via large-scale active time multiplexing,} {{Science
  Advances}} \textbf{5}, eaaw8586 (2019).

\bibitem{somaschi2016}
N.~Somaschi, V.~Giesz, L.~De~Santis, J.~Loredo, M.~P. Almeida, G.~Hornecker,
  S.~L. Portalupi, T.~Grange, C.~Ant{\'o}n, J.~Demory \emph{et~al.},
  \enquote{Near-optimal single-photon sources in the solid state,}
  {{Nature Photon}} \textbf{10}, 340--345 (2016).

\bibitem{loredo2016}
J.~C. Loredo, N.~A. Zakaria, N.~Somaschi, C.~Anton, L.~de~Santis, V.~Giesz,
  T.~Grange, M.~A. Broome, O.~Gazzano, G.~Coppola, I.~Sagnes, A.~Lemaitre,
  A.~Auffeves, P.~Senellart, M.~P. Almeida, and A.~G. White, \enquote{Scalable
  performance in solid-state single-photon sources,}
  {{Optica}} \textbf{3}, 433--440 (2016).

\bibitem{wang2017}
H.~Wang, Y.~He, Y.-H. Li, Z.-E. Su, B.~Li, H.-L. Huang, X.~Ding, M.-C. Chen, C.~Liu, J.~Qin \emph{et~al.}, \enquote{High-efficiency multiphoton boson sampling,} {{Nature Photon}} \textbf{11}, 361--365 (2017).

\bibitem{kir2017} 
G.~Kir\ifmmode \check{s}\else \v{s}\fi{}ansk\ifmmode~\dot{e}\else \.{e}\fi{},   H.~Thyrrestrup, R.~S. Daveau, C.~L. Dree\ss{}en, T.~Pregnolato, L.~Midolo, P.~Tighineanu, A.~Javadi, S.~Stobbe, R.~Schott, A.~Ludwig, A.~D. Wieck, S.~I. Park, J.~D. Song, A.~V. Kuhlmann, I.~S\"ollner, M.~C. L\"obl, R.~J. Warburton, and P.~Lodahl, \enquote{Indistinguishable and efficient single photons from a quantum dot in a planar nanobeam waveguide,} {{Phys. Rev. B}} \textbf{96}, 165306 (2017).

\bibitem{wang2019p}
H.~Wang, Y.-M. He, T.-H. Chung, H.~Hu, Y.~Yu, S.~Chen, X.~Ding, M.-C. Chen, J.~Qin, X.~Yang \emph{et~al.}, \enquote{Towards optimal single-photon sources from polarized microcavities,} {{Nature Photon}} \textbf{13}, 770--775 (2019).

\bibitem{maunz2007} 
P.~Maunz, D.~Moehring, S.~Olmschenk, K.~Younge, D.~Matsukevich, and C.~Monroe, \enquote{Quantum interference of photon pairs from two remote trapped atomic ions,} {{Nature Phys}} \textbf{3}, 538--541 (2007).

\bibitem{rosenfeld2017}
W.~Rosenfeld, D.~Burchardt, R.~Garthoff, K.~Redeker, N.~Ortegel, M.~Rau, and
  H.~Weinfurter, \enquote{Event-ready bell test using entangled atoms
  simultaneously closing detection and locality loopholes,}
  {{Phys. Rev. Lett.}} \textbf{119}, 010402 (2017).

\bibitem{thompson2006}
J.~K. Thompson, J.~Simon, H.~Loh, and V.~Vuleti{\'c}, \enquote{A
  high-brightness source of narrowband, identical-photon pairs,}
  {{Science}} \textbf{313}, 74--77 (2006).

\bibitem{wilk2007}
T.~Wilk, S.~C. Webster, H.~P. Specht, G.~Rempe, and A.~Kuhn,
  \enquote{Polarization-controlled single photons,}
  {{Phys. Rev. Lett.}} \textbf{98}, 063601 (2007).

\bibitem{nisbet2011}
P.~B.~R. Nisbet-Jones, J.~Dilley, D.~Ljunggren, and A.~Kuhn, \enquote{Highly efficient source for indistinguishable single photons of controlled shape,} {{New Journal of Physics}} \textbf{13}, 103036 (2011).

\bibitem{mucke2013} 
M.~M\"ucke, J.~Bochmann, C.~Hahn, A.~Neuzner, C.~N\"olleke, A.~Reiserer, G.~Rempe, and S.~Ritter, 
\enquote{Generation of single photons from an atom-cavity system,} {{Phys. Rev. A}} \textbf{87}, 063805 (2013).

\bibitem{clark2019}
L.~W. Clark, N.~Jia, N.~Schine, C.~Baum, A.~Georgakopoulos, and J.~Simon, 
\enquote{Interacting floquet polaritons,} {{Nature}} \textbf{571}, 532--536 (2019).



\end{thebibliography}

\begin{thebibliography}{10}
\newcommand{\enquote}[1]{``#1''}

\bibitem{srosi2018}
S.~Rosi, A.~Burchianti, S.~Conclave, D.~S. Naik, G.~Roati, C.~Fort, and
  F.~Minardi, \enquote{$\lambda$-enhanced grey molasses on the $d_2$ transition
  of rubidium-87 atoms,} {{Scientific reports}}
  \textbf{8}, 1301 (2018).

\bibitem{sleseluc2018}
S.~de~L\'es\'eleuc, D.~Barredo, V.~Lienhard, A.~Browaeys, and T.~Lahaye,
  \enquote{Analysis of imperfections in the coherent optical excitation of
  single atoms to rydberg states,} {{Phys. Rev. A}}
  \textbf{97}, 053803 (2018).

\bibitem{sARC}
N.~{\v{S}}ibali{\'c}, J.~D. Pritchard, C.~S. Adams, and K.~J. Weatherill,
  \enquote{Arc: An open-source library for calculating properties of alkali
  rydberg atoms,} {{Computer Physics Communications}}
  \textbf{220}, 319--331 (2017).

\bibitem{ssaffman2002}
M.~Saffman and T.~G. Walker, \enquote{Creating single-atom and single-photon
  sources from entangled atomic ensembles,} {{Phys. Rev.
  A}} \textbf{66}, 065403 (2002).

\bibitem{sStevens2013}
M.~J. Stevens, \enquote{{Photon Statistics, Measurements, and Measurements
  Tools},} in \emph{Experimental Methods in the Physical Sciences,}  (Elsevier,
  2013), pp. 25--52.

\bibitem{suppu2016}
R.~Uppu, T.~A. Wolterink, T.~B. Tentrup, and P.~W. Pinkse, \enquote{Quantum
  optics of lossy asymmetric beam splitters,} {{Optics
  express}} \textbf{24}, 16440--16449 (2016).

\bibitem{squtip}
J.~R. Johansson, P.~D. Nation, and F.~Nori, \enquote{Qutip 2: A python
  framework for the dynamics of open quantum systems,}
  {{Computer Physics Communications}} \textbf{184},
  1234--1240 (2013).

\bibitem{salexey2007}
A.~V. Gorshkov, A.~Andr\'e, M.~D. Lukin, and A.~S. S\o{}rensen, \enquote{Photon
  storage in $\ensuremath{\Lambda}$-type optically dense atomic media. ii.
  free-space model,} {{Phys. Rev. A}} \textbf{76}, 033805
  (2007).

\bibitem{swang2016}
X.-L. Wang, L.-K. Chen, W.~Li, H.-L. Huang, C.~Liu, C.~Chen, Y.-H. Luo, Z.-E.
  Su, D.~Wu, Z.-D. Li, H.~Lu, Y.~Hu, X.~Jiang, C.-Z. Peng, L.~Li, N.-L. Liu,
  Y.-A. Chen, C.-Y. Lu, and J.-W. Pan, \enquote{Experimental ten-photon
  entanglement,} {{Phys. Rev. Lett.}} \textbf{117}, 210502
  (2016).

\bibitem{sxiong2016}
C.~Xiong, X.~Zhang, Z.~Liu, M.~J. Collins, A.~Mahendra, L.~Helt, M.~J. Steel,
  D.-Y. Choi, C.~Chae, P.~Leong \emph{et~al.}, \enquote{Active temporal
  multiplexing of indistinguishable heralded single photons,}
  {{Nat Commun}} \textbf{7}, 10853 (2016).

\bibitem{skaneda2019}
F.~Kaneda and P.~G. Kwiat, \enquote{High-efficiency single-photon generation
  via large-scale active time multiplexing,} {{Science
  Advances}} \textbf{5}, eaaw8586 (2019).

\bibitem{ssomaschi2016}
N.~Somaschi, V.~Giesz, L.~De~Santis, J.~Loredo, M.~P. Almeida, G.~Hornecker,
  S.~L. Portalupi, T.~Grange, C.~Ant{\'o}n, J.~Demory \emph{et~al.},
  \enquote{Near-optimal single-photon sources in the solid state,}
  {{Nature Photon}} \textbf{10}, 340--345 (2016).

\bibitem{sloredo2016}
J.~C. Loredo, N.~A. Zakaria, N.~Somaschi, C.~Anton, L.~de~Santis, V.~Giesz,
  T.~Grange, M.~A. Broome, O.~Gazzano, G.~Coppola, I.~Sagnes, A.~Lemaitre,
  A.~Auffeves, P.~Senellart, M.~P. Almeida, and A.~G. White, \enquote{Scalable
  performance in solid-state single-photon sources,}
  {{Optica}} \textbf{3}, 433--440 (2016).

\bibitem{swang2017}
H.~Wang, Y.~He, Y.-H. Li, Z.-E. Su, B.~Li, H.-L. Huang, X.~Ding, M.-C. Chen,
  C.~Liu, J.~Qin \emph{et~al.}, \enquote{High-efficiency multiphoton boson
  sampling,} {{Nature Photon}} \textbf{11}, 361--365
  (2017).

\bibitem{skir2017}
G.~Kir\ifmmode \check{s}\else \v{s}\fi{}ansk\ifmmode~\dot{e}\else \.{e}\fi{},
  H.~Thyrrestrup, R.~S. Daveau, C.~L. Dree\ss{}en, T.~Pregnolato, L.~Midolo,
  P.~Tighineanu, A.~Javadi, S.~Stobbe, R.~Schott, A.~Ludwig, A.~D. Wieck, S.~I.
  Park, J.~D. Song, A.~V. Kuhlmann, I.~S\"ollner, M.~C. L\"obl, R.~J.
  Warburton, and P.~Lodahl, \enquote{Indistinguishable and efficient single
  photons from a quantum dot in a planar nanobeam waveguide,}
  {{Phys. Rev. B}} \textbf{96}, 165306 (2017).

\bibitem{swang2019p}
H.~Wang, Y.-M. He, T.-H. Chung, H.~Hu, Y.~Yu, S.~Chen, X.~Ding, M.-C. Chen,
  J.~Qin, X.~Yang \emph{et~al.}, \enquote{Towards optimal single-photon sources
  from polarized microcavities,} {{Nature Photon}}
  \textbf{13}, 770--775 (2019).

\bibitem{smaunz2007}
P.~Maunz, D.~Moehring, S.~Olmschenk, K.~Younge, D.~Matsukevich, and C.~Monroe,
  \enquote{Quantum interference of photon pairs from two remote trapped atomic
  ions,} {{Nature Phys}} \textbf{3}, 538--541 (2007).

\bibitem{srosenfeld2017}
W.~Rosenfeld, D.~Burchardt, R.~Garthoff, K.~Redeker, N.~Ortegel, M.~Rau, and
  H.~Weinfurter, \enquote{Event-ready bell test using entangled atoms
  simultaneously closing detection and locality loopholes,}
  {{Phys. Rev. Lett.}} \textbf{119}, 010402 (2017).

\bibitem{sthompson2006}
J.~K. Thompson, J.~Simon, H.~Loh, and V.~Vuleti{\'c}, \enquote{A
  high-brightness source of narrowband, identical-photon pairs,}
  {{Science}} \textbf{313}, 74--77 (2006).

\bibitem{swilk2007}
T.~Wilk, S.~C. Webster, H.~P. Specht, G.~Rempe, and A.~Kuhn,
  \enquote{Polarization-controlled single photons,}
  {{Phys. Rev. Lett.}} \textbf{98}, 063601 (2007).

\bibitem{snisbet2011}
P.~B.~R. Nisbet-Jones, J.~Dilley, D.~Ljunggren, and A.~Kuhn, \enquote{Highly
  efficient source for indistinguishable single photons of controlled shape,}
  {{New Journal of Physics}} \textbf{13}, 103036 (2011).

\bibitem{smucke2013}
M.~M\"ucke, J.~Bochmann, C.~Hahn, A.~Neuzner, C.~N\"olleke, A.~Reiserer,
  G.~Rempe, and S.~Ritter, \enquote{Generation of single photons from an
  atom-cavity system,} {{Phys. Rev. A}} \textbf{87},
  063805 (2013).

\end{thebibliography}
\end{document}